\RequirePackage{fix-cm}
\documentclass{svjour3}
\smartqed

\DeclareUnicodeCharacter{2212}{-}

\usepackage{adjustbox}
\usepackage[utf8]{inputenc}
\usepackage{acronym}
\usepackage{graphicx}
\usepackage[T1]{fontenc}
\usepackage{epsfig}
\usepackage{txfonts}
\usepackage{amssymb}
\usepackage{siunitx}
\usepackage[backend=biber,sorting=none,minnames=3,defernumbers=true]{biblatex}
\usepackage{tocloft}
\usepackage{hyperref}
\usepackage{rotating}
\usepackage{pdflscape}
\usepackage{afterpage}

\journalname{ISSI Scientific Reports}

\begin{document}

\title{ESA Science Programme Missions: Contributions and Exploitation -- INTEGRAL Observing Time Proposals}

\titlerunning{INTEGRAL Observing Proposals}
\authorrunning{Kuulkers et al.}

\author{Erik Kuulkers \and Celia S\'anchez-Fern\'andez \and Arvind N. Parmar}

\institute{E. Kuulkers \at
Directorate of Science, ESA/ESTEC\\
Noordwijk, The Netherlands.\\
\email{Erik.Kuulkers@esa.int}
\and
C. S\'anchez-Fern\'andez \at
Directorate of Science, ESA/ESAC\\ 
Villanueva de la Ca\~nada, Madrid, Spain.\\
\and 
A. Parmar \at
Former Head of the Science Support Office\\
Directorate of Science, ESA/ESTEC\\
The Netherlands\\
\emph{Present address:}\\
Department of Space and Climate Physics\\
MSSL/UCL\\
Dorking\\ 
UK\\
}
                      
\date{Received:  date /  Accepted:  date}

\maketitle

\begin{abstract}
    We examine the outcomes of the regular announcements of observing opportunities for ESA's gamma-ray observatory INTEGRAL issued between 2000 and 2021. We investigate how success rates vary with the lead proposer's gender, "academic age" and the country where the proposer's institute is located. The more than 20 years operational lifetime enable the evolution of the community proposing for INTEGRAL to be probed. We determine proposal success rates for high-priority and all proposals using both the numbers of accepted proposals and the amounts of awarded observing time. We find that male lead proposers are between 2--11\% more successful than their female counterparts in obtaining INTEGRAL observations. We investigate potential correlations between the female-led proposal success rates and the amount of female participation in the Time Allocation Committee. 
\end{abstract}  

\section{\textbf{Introduction}}

\ac{INTEGRAL} was launched from the Baikonur cosmodrome on 17 October 2002 into a 72-hour highly elliptical orbit by a Proton rocket.  The nominal mission duration was 2 years with resources available for a longer extended mission. In January 2015 the orbit was changed to a 64~hour orbital period to ensure the safe disposal of the mission in 2029. This ESA-led mission includes contributions from Russia (launcher) and \ac{NASA} (Goldstone ground station).  \ac{INTEGRAL} provides an unprecedented combination of celestial imaging and spectroscopy over a wide range of X-ray and gamma-ray energies as well as simultaneous optical monitoring \cite{2003Winkler}, \cite{2021Kuulkers}. Many details about the INTEGRAL spacecraft, orbit, instruments, scientific aims and first results can be found in volume 411 of A\&A (2003), a special Astronomy \&\ Astrophysics issue dedicated to INTEGRAL. The \ac{INTEGRAL} payload consists of two gamma-ray instruments, one of which is optimised for 15~\ac{keV} to 10~\ac{MeV} high-resolution imaging (\ac{IBIS};~\cite{2003Ubertini}) and the other for 20~keV to 8~\ac{MeV} high-resolution spectroscopy (\ac{SPI};~\cite{2003Vedrenne}). \ac{IBIS} provides an angular resolution of 12$'$ \ac{FWHM} over a  30$\times$30~degrees$^2$ \ac{FOV} and an energy resolution, E/$\Delta$E, of $\sim$12~\ac{FWHM} at 100~\ac{keV}. \ac{SPI} provides an angular resolution of 2.7 degrees \ac{FWHM} over a 30$\times$30~degrees$^2$ \ac{FOV} and an E/$\Delta$E of $\sim$500~FWHM at 1.3~\ac{MeV}. Both instruments provide millisecond time resolution. The extremely broad energy range of \ac{IBIS} is covered by two separate detector arrays, \ac{ISGRI} (15--500~\ac{keV}) and \ac{PICsIT} (0.2--10~\ac{MeV}). The payload is completed by two \ac{JEM-X}; 3--35~\ac{keV};~\cite{2003Lund}) and optical (\ac{OMC}; V-band; ~\cite{2003MasHesse}) monitors.  The instruments are co-aligned and are normally operated simultaneously. By using the anti-coincidence shields of \ac{IBIS} and \ac{SPI} as active detectors, an omni-directional sensitivity above 75~\ac{keV} is achieved. The highly eccentric orbit (period 2.7-days) allows the full sky to be continuously monitored for about 85\%\ of the time (when INTEGRAL is above the Earth’s radiation belts). The fraction of sky occulted by the Earth is much smaller than for satellites in low-Earth orbits.

The \ac{MOC} is located at \ac{ESOC}, in Darmstadt, Germany. The \ac{SOC} was originally in \ac{ESTEC}, Noordwijk, The Netherlands but moved to \ac{ESAC}, near Madrid, Spain in February 2005. The \ac{SOC} receives observation proposals and optimises the accepted ones into an observation plan consisting of a time line of target pointings together with the corresponding instrument configurations. The \ac{ISDC} receives the science telemetry plus the relevant ancillary spacecraft data from the \ac{MOC} which is responsible for the operations of the spacecraft and payload. The \ac{ISDC} Data Centre for Astrophysics~\cite{2003Courvoiser} processes the received telemetry and generates standard data products which are distributed and archived. The archive at the ISDC receives about 300 independent visits per month, a rate that has been stable over the past decade. INTEGRAL data can be processed using the \ac{OSA} software provided by the \ac{ISDC}.  This includes pipelines for the reduction of INTEGRAL data from all four instruments. The three high-energy instruments use coded masks to provide imaging information. This means that photons from a source within the \ac{FOV} are distributed over the detector area in a pattern determined by the position of the source in the \ac{FOV}. Source positions and intensities are determined by matching the observed distribution of counts with those produced by the mask modulation. The energy range and use of coded masks means that observation lengths are usually longer than for directly imaging instruments due to the higher background in the source extraction regions.

\section{\textbf{Science Results}}

INTEGRAL has become part of the newly-established multi-messenger astrophysics science network, measuring gamma-rays coincident with \ac{GW}, ultra-high energy neutrino signals and \ac{FRB}s, so linking the new non-electromagnetic domains with the hard X- and gamma-ray electromagnetic radiation \cite{2017NatAs...1E..83V}, \cite{2020ApJ...898L..29M}, \cite{2021NewAR..9201595F}. This synergy is achieved primarily by exploiting two of INTEGRAL’s key capabilities -- the highly efficient coverage of nearly the whole sky and the rapid reaction capability for unexpected events. INTEGRAL can rapidly (typically within half a day) re-point and conduct \ac{ToO} observations. Of all the \ac{ToO} observations performed until the end of 2021, 4\%\ started within 8 hours, while 85\%\ started 3 days or later after the alert. 

A key part of INTEGRAL's legacy is the continuing discovery of new high-energy sources, now amounting to more than 700 \cite{2016ApJS..223...15B}, \cite{2022MNRAS.510.4796K}. Using multi-wavelength follow-up observations, about 250 of the newly discovered sources have been firmly identified, and their dominant emission mechanisms recognised. The classes of objects discovered include Seyfert galaxies, blazars, cataclysmic variables, supergiant fast X-ray transients. INTEGRAL also provides unique measurements of the decay lines of radioactive isotopes, e.g., from extragalactic supernovae and for the study of stellar feedback in our Galaxy. The extent and morphology of the enigmatic positron annihilation in the central regions of the Galaxy, possibly linked to dark-matter decay, as well as of $^{26}$Al and $^{60}$Fe continues to be mapped by INTEGRAL \cite{2016A&A...586A..84S}, \cite{2020ApJ...889..169W}, \cite{2022MNRAS.509L..11S}. INTEGRAL is also performing pioneering and unique gamma-ray polarization studies of \ac{GRB}s and bright Galactic sources, which are fundamental in understanding the emission mechanisms.  The rich science which has resulted from INTEGRAL’s observations over its lifetime are presented in the various review articles in the special issue of New Astronomy Reviews "Fifteen-plus years of INTEGRAL science" available at \url{https://www.sciencedirect.com/journal/new-astronomy-reviews/special-issue/10K9PV78WN4}.
Here we provide some examples to illustrate the scientific impact of INTEGRAL.

{\bf Gravitational wave sources:} The INTEGRAL and Fermi~\cite{2021Ajello} detection of the short \ac{GRB} (GRB 170817A) linked to a \ac{GW} event (GW170817) on 17 August 2017 caused by the merger of a binary neutron star \cite{2017ApJ...848L..13A}, \cite{2017ApJ...848L..15S}, was a fundamental milestone for the wider astrophysics and physics communities. Using INTEGRAL, the Fermi localisation area of the \ac{GRB} could be improved by more than a factor two; this has been important to prove, not only the temporal, but also the spatial coincidence of the gamma-ray emission and the \ac{GW} event. INTEGRAL has provided instantaneous and complete coverage of the large localisation regions of most of the other \ac{GW} events detected by the LIGO/Virgo collaboration. The majority of them have been associated with binary black holes, for which most models do not predict detectable electromagnetic signals: whenever INTEGRAL was above the radiation belts, it provided strict upper limits.

{\bf Bright transients:} Several bright black-hole binary transients have led to unique, extensive \ac{ToO} campaigns with INTEGRAL. In 2015, V404 Cygni, one of the closest (2.4~\ac{kpc}) and best-established accreting black-hole (about 9~M$_{\odot}$) binaries, became the brightest object in the X-ray sky (up to 50 times the flux of the Crab). Its outburst showed highly unusual behaviour with repeated bright optical and X-rays flashes on time scales shorter than one hour. It was observed almost continuously by INTEGRAL simultaneously in the optical and hard X-ray bands, which was not possible with any other mission. More than 40 refereed publications have been written using these data. Transients of this type can provide unique diagnostics of the physical processes operating near black holes through the study of optical and hard X-ray to gamma-ray emission \cite{2015A&A...581L...9R}, \cite{2015ApJ...813L..22R}, \cite{2015ApJ...813L..21N}, \cite{2017ApJ...834..130J}, \cite{2017ApJ...848....3R}, \cite{2017A&A...602A..40S}, \cite{2018A&A...616A.129K}, \cite{2019ApJ...870...92R}, \cite{2020A&A...634A..94K}, \cite{2021A&A...645A..60C}, \cite{2021ApJ...910...21R}, including positron annihilation signals \cite{1991SvAL...17..437G}, \cite{1992ApJ...389L..79G}, \cite{2016Natur.531..341S}, \cite{2019ApJ...870...92R}, and gamma-ray polarisation \cite{2017PoS.INTEGRAL2016.022}. The sensitive study of Comptonisation, pair emission, and polarisation above $\sim$100~\ac{keV} are unique strengths of INTEGRAL. 

{\bf Supernovae:} INTEGRAL discovered gamma-ray emission from the decay of $^{56}$Ni and $^{56}$Co in a Type Ia supernova, SN2014J. While confirming theoretical expectations these measurements significantly added new constraints on how $^{56}$Ni ejecta are spatially and kinematically distributed throughout an expanding envelope \cite{2014Natur.512..406C}, \cite{2014Sci...345.1162D}, \cite{2015A&A...574A..72D}, \cite{2015AN....336..464D}, \cite{2015ApJ...812...62C}, \cite{2016A&A...588A..67I}. For core-collapse supernovae, INTEGRAL spectroscopy of lines from the decay of $^{44}$Ti likewise provided essential kinematic information from the deep interiors of such supernovae (Cas A; \cite{2006ApJ...647L..41R}, \cite{2015A&A...579A.124S}), which are otherwise inaccessible. Both families of explosions have now been recognised to be very diverse. Nuclear, as well as non-thermal, emission occurs within INTEGRAL's energy range allowing for essential follow-up observations. 

{\bf Fast radio bursts:} One of the currently intriguing mysteries is the origin of millisecond duration, 10$^{40}$~erg, powerful \ac{FRB}s. The majority of \ac{FRB} do not repeat, but a growing number of newly discovered events (now $>$10) are repeaters \cite{2016Natur.531..202S}, \cite{2020Natur.582..351C}, \cite{2020ApJ...891L...6F}. In FRB180916.J0158+65 a 16-day periodicity suggests an origin from a compact object either in a binary system, or in an isolated, precessing, magnetar \cite{2020Natur.582..351C}. INTEGRAL is well suited to study such systems with typical emission at high energies \cite{2020ApJ...896L..40P}, \cite{2019ATel13073....1G}.  The near all-sky monitoring capability of INTEGRAL allowed a magnetar flare associated with a Galactic \ac{FRB} in the soft gamma-ray repeater SGR~1935+2154 to be found, showing that some \ac{FRB}s are associated with magnetars \cite{2020ApJ...898L..29M}. Since \ac{FRB}s are extragalactic, this event provides a "missing link" to a population of extragalactic magnetars. GRB200415A, a giant flare from a magnetar in NGC 253 \cite{2021Natur.589..211S} may be such a source.

{\bf Other transients:} \ac{ToO} observations based on multi-wavelength studies are important for understanding the diversity of transient events. A recent example is AT2018cow (a supernova variant), also referred to as a fast-rising blue optical transient, which revealed its internal engine as a temporary hard X-ray spectral component detected by \ac{IBIS} up to about 100~\ac{keV}, in addition to luminous and highly variable soft X-rays \cite{2019ApJ...872...18M}. Another case of a rapidly variable and rising nuclear transient with soft gamma-ray emission was AT2019pev, which may be evidence of a new class of flares from accreting supermassive black holes \cite{2019NatAs...3..242T}, \cite{2019ATel13170....1F}. This opens a still unexplored window bridging optical transients to explosive events, never observed before in this phase.

{\bf Serendipitous science:} The \ac{IREM} and a few of the Radiation-sensitive Field-Effect Transistors (RadFETs) on \ac{IBIS} are providing important information about the Earth’s local environment and contributing to interplanetary space weather \cite{2015ITNS...62.2784C}, \cite{2017SpWea..15..917M}. \ac{IREM} is also used as a calibrator of similar radiation monitors onboard other spacecrafts \cite{2018APh....98...28A}. Another area where INTEGRAL provides serendipitous science includes Solar System research, with the study of Earth’s aurorae and Solar flares \cite{2015JPhCS.632a2081S}, \cite{2018JSWSC...8A..40G}.

\section{\textbf{Scientific Productivity}}

Up to the end of 2021, there have been close to 2000 refereed publications from \ac{INTEGRAL} that make direct use of data from the mission including from its primary catalogues, or make quantitative predictions of results from the mission or describe \ac{INTEGRAL}, its instruments, operations, software or calibrations.
The scientific productivity of INTEGRAL is screened monthly by the INTEGRAL SOC, following an approach consistent with those used by the XMM-Newton \cite{2014Ness}, Chandra \cite{2012Rots}  and \ac{HST} \cite{2010Apai} projects. The resulting data are stored in a dedicated database from which information can be extracted when needed. 
The INTEGRAL SOC determines the \ac{ObsID} and instruments used to provide the scientific results reported in each publication. The \ac{ObsID}s are then used to access the INTEGRAL Science Legacy Archive (ISLA) to provide detailed information on the observations and proposal leading to relevant data set.
The information obtained is then combined to allow the scientific productivity of the mission to be investigated.
The major conclusions of this work are the following: 
\begin{itemize}
\item An average of 51 scientists per year become lead authors for the first time
on a refereed paper which directly uses INTEGRAL data.
\item Each refereed INTEGRAL publication receives an average of around five citations per year with a
long-term citation rate of two citations per year, more than five years after publication. 
\item About eighty percent of the articles citing INTEGRAL articles are not primarily INTEGRAL observational papers. 
\item The distribution of elapsed time between the execution of the latest observation used in a publication and the date of that publication peaks at about 2.5 years. The publication of observations taken under the \ac{ToO} program (see below) shows a similar distribution. In other words, no significant differences are seen by the publication of INTEGRAL Open Time and ToO observations. 
\item At least 95\%\ of the science observing time is used in at least one publication after 7 elapsed years.
\item The scientific productivity of INTEGRAL measured by the publication rate, number of new authors and
citation rate, has fluctuated over the time, and experienced a significant boost in 2017, following the INTEGRAL contribution to the detection of the electro-magnetic counterpart to GW170817. 
\end{itemize}

\section{\textbf{Observing Time}}

\ac{INTEGRAL} is an observatory executing observations mainly selected through peer review. Usually there are annual \ac{AO}s, see Table~\ref{tab:INTEGRAL_AO}. The first announcement (AO-1) was issued in November 2000, prior to launch, and received 277 proposals with a very high over-subscription in time of a factor 19.7. AO-2 was issued in July 2003 and covered observations for a one year interval. Due to the impending move of the \ac{ISOC} from \ac{ESTEC} to \ac{ESAC} in 2005, AO-3 covered an interval of 18 months and was issued in September 2004. Subsequent \ac{AO}s, except for AO-6 and AO-7, covered intervals of one year and were issued in the Spring of each year. AO-6 was initially foreseen to cover 1 year, but in November 2008 it was decided to extend the cycle by two months. AO-7 also covered an interval of 14 months. The following types of observing time are available:

\begin{itemize}
\item{} Normal observations: observations which do not have special scheduling requirements allowing for the most efficient scheduling. Observation durations are $<$1 \ac{Msec}.
\item{} Key Programme observations: Observations intended to carry out scientific investigations requiring a significant amount of the total non-Target-of-Opportunity (non-\ac{ToO}) observing time during an \ac{AO} cycle ($>$1 Msec), but also accommodating various scientific aims. These cannot be \ac{ToO} observations.
\item{} Fixed Time observations: observations with special scheduling requirements such as phase-dependent observations of a binary system, coordinated multi-wavelength observations, or a sequence of observations separated by a time interval. Such observations usually reduce scheduling efficiency, because the spacecraft must be pointing towards a particular source at a particular time.
\item{} \ac{ToO} observations that have critical scheduling requirements and are meant as a fast response to new phenomena, such as X-ray novae outbursts, flares from active galactic nuclei, supernovae, and bright states of galactic micro-quasars. \ac{ToO}s can be for known or unknown sources identified by their class.
\item{} Multi-year proposals spanning two annual AOs for "Key Programmes" requiring substantial amounts of observing time that cannot be accommodated in a single AO.
\item{} Joint observations with the XMM-Newton, Swift, NuSTAR, and Fermi missions.
\item{} Guaranteed Time: from 2002 to 2008 the observing programme was divided into Open Time for the general observer (\ac{GP}) and the \ac{GT} for the \ac{ISWT} (\ac{CP}). The \ac{GT} during the \ac{CP} was the return to the \ac{ISWT} for their contributions to the development and execution of the INTEGRAL project. The \ac{CP} decreased from 35\%\ in the first year to 20\%\ in the sixth year after launch; from 2009 there was no further \ac{CP} (see Table~\ref{tab:INTEGRAL_GT}).
From AO-1 to AO-6 27\% of the Open Time was reserved for scientists from the Russian Federation in return for the provision of the launcher. From AO-7 this reduced to 25\%. The \ac{TAC} awards this guaranteed return on the same basis of scientific merit as all other proposals. 
\item{} Discretionary Time: The INTEGRAL Project Scientist can grant "Discretionary Time" which is $\sim$5\% of the available observing time. This can be used for unanticipated \ac{ToO}s -- observations where a timely observation outside the normal AO cycle is likely to result in a significant scientific impact. Discretionary Time observations may have a proprietary period of six months, or may be made publicly available as soon as the relevant data files have been created.
\item{} Around 5\% of the available observing time is used to maintain the calibration and monitor instrument health.
\end{itemize} 

The detailed contents of the AOs varied. In AO-4 there was a pilot Key Programme for proposals asking for $>$1 \ac{Msec} of observing time. In AO-5 and AO-6 there were dedicated calls for Key Programmes which could continue for more than one AO. In AO-7 to AO-19 Key Programme proposals which lasted for up to 2 years could be submitted. If accepted, these had to be resubmitted the following year. In AO-5 Key Programme proposals accepted for more than one year were automatically taken over to the next year.  From AO-4 to AO-11 there were separate calls for data rights proposals. These were for targets in the large \ac{IBIS} and \ac{SPI} \ac{FOV}s that were observed "serendipitously" in other proposals. Accepted proposals were awarded zero observing time but had the right to analyse data from the approved target, or targets. A few of the proposers in AO-1 also asked for serendipitous time in other programmes, and these were also treated as data rights proposals. 
From AO-12 to AO-19 the two types of proposals were merged, and data right proposers had to request time, but once accepted, they were awarded zero time. 
Data from fast transient events (such as \ac{GRB}s) may be contained in the science data of the operating instruments. Data on such an event itself can be asked for, and are handled, in principle, as a \ac{ToO} event: if there are \ac{TAC}-approved proposals for such events in the \ac{FOV} and it is detected during an INTEGRAL observation, then the successful proposer will be granted the data rights on the event. Again, these are serendipitous events, and thus awarded zero time.
If the positions of accepted targets are close-by, proposals could be merged into one observation with an optimised exposure time (called "amalgamation"), in order to save slewing time and to maximise the observation efficiency. Those proposals which were amalgamated to the main proposal were treated as obtaining zero time. In this chapter we only take into account Open Time proposals.

\begin{figure*}[ht]
\centering
\includegraphics[width=1.0\textwidth,angle=0]{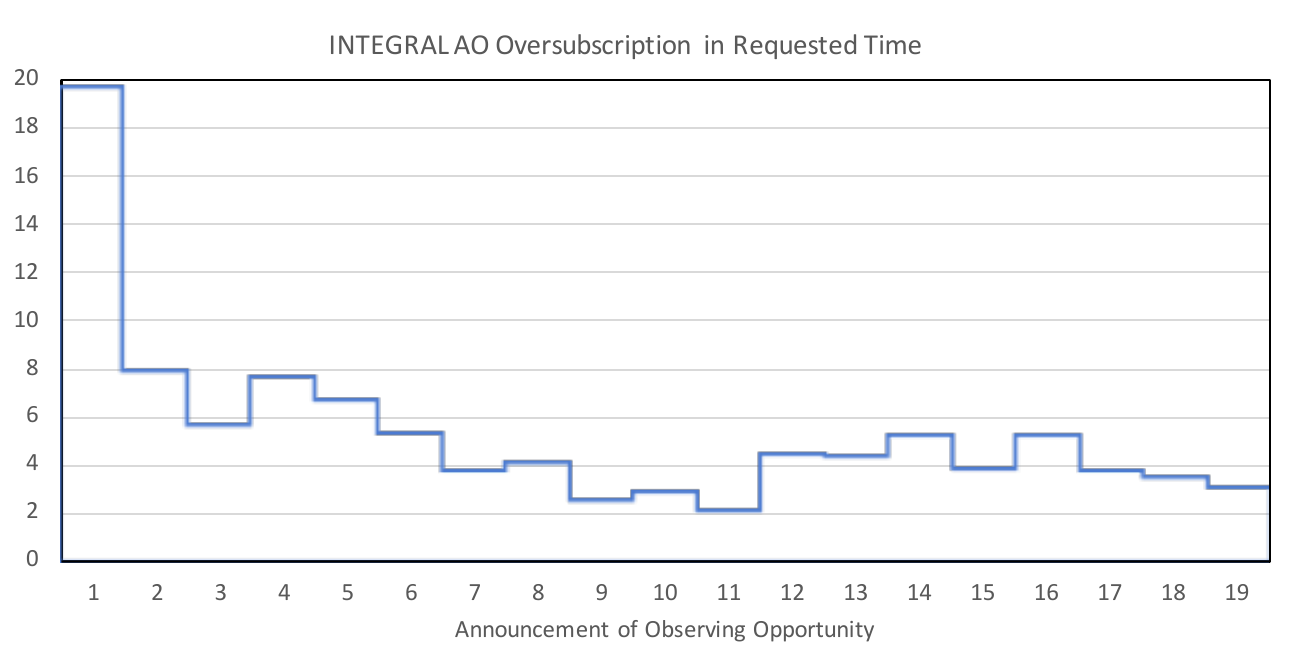}
\caption{The over-subscription of requested INTEGRAL observing time compared to that available for AO-1 to AO-19, covering an interval of 21 years from 2000.}
\label{fig:INTEGRAL AO}    
\end{figure*}

\begin{table}
\centering
\caption{INTEGRAL \ac{AO} summary. The numbers of \ac{TAC} members include the panel chairs.}
\setlength\tabcolsep{2.6pt}
\begin{tabular}{lcccccccc}
\hline \noalign{\smallskip}
\vspace{3pt}
 AO  & Year      &  Observation           & Total No. of        &    Over-subs.        & \multicolumn{4}{c}{TAC}\\
 & Issued & Interval & Proposals  & (Time)   & \multicolumn{2}{c}{Members} & \multicolumn{2}{c}{Panel Chairs} \\
 & & &  &           &   Male &    Female & Male & Female \\
\hline \noalign{\smallskip}
 1 & 2000 & 30 Dec 2002 -- 16 Dec 2003 & 277 & 19.7 & 22 & 5 & 4 & 0 \\
 2 & 2003 & 17 Dec 2003 -- 17 Feb 2005 & 151 &  7.9 & 21 & 6 & 4 & 0 \\
 3 & 2004 & 18 Feb 2005 -- 15 Aug 2006 & 112 &  5.7 & 20 & 6 & 3 & 0 \\
 4 & 2006 & 16 Aug 2006 -- 16 Aug 2007 & 145 &  7.7 & 15 & 4 & 3 & 0 \\
 5 & 2007 & 17 Aug 2007 -- 16 Aug 2008 & 183 &  6.7 & 17 & 2 & 3 & 0 \\
 6 & 2008 & 17 Aug 2008 -- 15 Oct 2009 & 184 &  5.3 & 19 & 0 & 3 & 0 \\
 7 & 2009 & 16 Oct 2009 -- 31 Dec 2010 & 157 &  3.8 & 13 & 3 & 3 & 0 \\
 8 & 2010 & 01 Jan 2011 -- 31 Dec 2011 & 132 &  4.1 & 12 & 4 & 3 & 0 \\
 9 & 2011 & 01 Jan 2012 -- 31 Dec 2012 & 101 &  2.6 & 13 & 3 & 2 & 1 \\
10 & 2012 & 01 Jan 2013 -- 31 Dec 2013 &  97 &  2.9 & 13 & 3 & 2 & 1 \\
11 & 2013 & 01 Jan 2014 -- 31 Dec 2014 &  87 &  2.1 & 14 & 2 & 1 & 2 \\
12 & 2014 & 01 Jan 2015 -- 31 Dec 2015 &  77 &  4.5 & 13 & 3 & 1 & 2 \\
13 & 2015 & 01 Jan 2016 -- 31 Dec 2016 &  63 &  4.4 & 12 & 4 & 3 & 0 \\
14 & 2016 & 01 Jan 2017 -- 31 Dec 2017 &  58 &  5.2 & 12 & 4 & 3 & 0 \\
15 & 2017 & 01 Jan 2018 -- 31 Dec 2018 &  65 &  3.9 & 11 & 5 & 2 & 1 \\
16 & 2018 & 01 Jan 2019 -- 31 Dec 2019 &  62 &  5.2 & 11 & 5 & 2 & 1 \\
17 & 2019 & 01 Jan 2020 -- 31 Dec 2020 &  63 &  3.8 & 12 & 4 & 1 & 2 \\
18 & 2020 & 01 Jan 2021 -- 31 Dec 2021 &  52 &  3.5 & 11 & 5 & 1 & 2 \\
19 & 2021 & 01 Jan 2022 -- 31 Dec 2022 &  49 &  3.1 & 11 & 5 & 2 & 1 \\ 
\textbf{Total} & & &\textbf{2115} & & \textbf{272} & \textbf{73} & \textbf{46} & \textbf{13} \\
\hline \noalign{\smallskip}
\end{tabular}
\label{tab:INTEGRAL_AO}
\end{table}

\begin{table}
\centering
\caption{INTEGRAL Guaranteed Time (GT) compared to the available observing time.}
\begin{tabular}{lcl}
\hline \noalign{\smallskip}
\vspace{3pt}
Period & GT (\%) & Mission Phase \\
\hline \noalign{\smallskip}
17 Dec 2002 -- 16 Dec 2003 & 35 & Nominal mission \\
17 Dec 2003 -- 16 Dec 2004 & 30 & Nominal mission \\
17 Dec 2004 -- 16 Dec 2005 & 25 & Extended mission \\
17 Dec 2005 -- 16 Dec 2006 & 25 & Extended mission \\
17 Dec 2006 -- 16 Dec 2007 & 25 & Extended mission \\
17 Dec 2007 -- 16 Dec 2008 & 20 & Extended mission \\
\hline \noalign{\smallskip}
\end{tabular}
\label{tab:INTEGRAL_GT}
\end{table}

\section{\textbf{Time Allocation Committee (TAC)}}

The \ac{TAC} reviews all submitted proposals and makes recommendations to the \ac{D-SCI} on the observing programme to be performed by INTEGRAL. It takes into account the scientific case, justification, merit and relevance of the proposed observation(s), and the potential contribution of the overall scientific return of the mission. The \ac{TAC} recommendations on the observing programme are given priority assignments of A, B or C, with A being the highest priority (in AO-1 and AO-2 accepted \ac{ToO} proposals did not receive a grade and we treat them as being grade A). Priority A and B targets are of major scientific importance and are scheduled with the highest priority. Priority C targets are used as "fillers" and have a significant lower likelihood to be finally scheduled. Priority A and B targets are automatically transferred to the next observation period, if their successful observation commenced (i.e., at least 25\% executed of the approved time), but is uncompleted during the current one. The total recommended observing time is about quarter more than the total available Open Time, for schedule planning efficiency.  Approximately 45\% of the recommended observing time is designated as Priority A, 30\% as Priority B and 25\% as Priority C.
For \ac{ToO} proposals, the actual requested observing time (other than the total time requested based on the number of targets) has been taken into account. 

An average of 110 valid proposals were received in response to each AO (see Table~\ref{tab:INTEGRAL_AO}). With the over-subscription\footnote{The over-subscription for INTEGRAL is defined as the ratio of the requested time over the available time. The available time is the total time expected to be available for Guest Observer observations in an upcoming AO observing period; for the later AOs this was about 21 Msec (note that this number does not include perigee passages, engineering observations, calibration observations, etc.). The requested time is the total time requested in non-ToO proposals plus 10\%\ of the total time asked in ToO proposals. The 10\%\ is a kind of rough average probability that a ToO proposal is triggered.} factors (see Fig.~\ref{fig:INTEGRAL AO}) scientific assessment of the proposals is a major undertaking of the astronomical community with over 340 scientists having participated so far. For the first two AOs, the \ac{TAC} was divided into four panels to reflect the range of topics proposed. This was reduced to three panels for subsequent AOs. Much of the assessment is performed by panels, consisting typically of five scientists selected from the worldwide community.  Each panel is led by a panel chair and the overall \ac{TAC} is led by a chairperson. The names of the \ac{TAC} members (with exception of the chairperson who sits on the \ac{IUG}, see Sect.~\ref{subsec:IUG}) are not made public.

\begin{figure*}[ht]
\centering
\includegraphics[width=1.0\textwidth,angle=0]{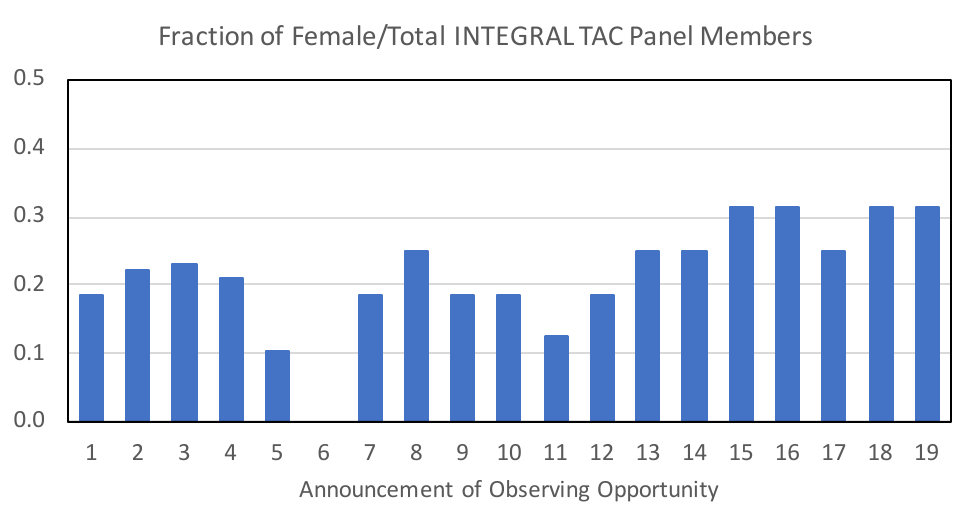}
\caption{The fraction of female INTEGRAL \ac{TAC} members compared to the total for AO-1 to AO-19 covering an interval of around 21 years. An increasing trend in the fraction of females \ac{TAC} members from $\sim$0.2 in AO-1 to $\sim$0.3 by AO-19 is evident.}
\label{fig:INTEGRAL TAC Members}    
\end{figure*}

\begin{figure*}[ht]
\centering
\includegraphics[width=1.0\textwidth,angle=0]{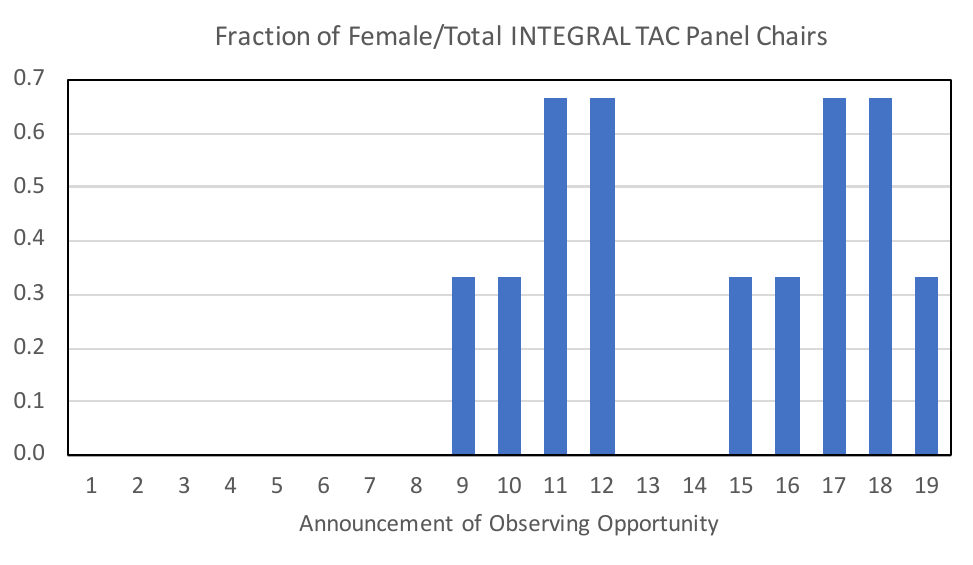}
\caption{The fraction of female INTEGRAL \ac{TAC} chairs compared to the total for AO-1 to AO-19 covering an interval of around 21 years. We note that there were no female panel chairs until AO-9 in 2011.}
\label{fig:INTEGRAL TAC Chairs}    
\end{figure*}

The genders of the \ac{TAC} members and panel chairs were assigned through the personal knowledge of the authors and \ac{SOC} staff. We appreciate that gender identity is more complex than a binary issue. However, no attempt was made to assign genders other than male or female in this study.  We then examined the gender composition of the \ac{TAC} members and panel chairs (Table~\ref{tab:INTEGRAL_AO}) from AO-1 to AO-19 covering 2001 to 2021. \ac{TAC} Panel members are chosen for their high-level of relevant scientific knowledge, while selection of the \ac{TAC} chairs was for scientists considered to have leadership roles in high-energy astronomy. Fig.~\ref{fig:INTEGRAL TAC Members} shows the fraction of female INTEGRAL \ac{TAC} members and Fig.~\ref{fig:INTEGRAL TAC Chairs} the same for the panel chairs. In total there were 226 male and 60 female panel members of which 46 males and 13 females were panel chairs. Since members and chairs may serve for multiple AOs, the numbers of individuals involved is smaller. Most of the \ac{TAC} members (and chairs) are based at institutes located in the \ac{ESA} Member States, with a significant number from institutes in the Russian Federation\footnote{By agreement, 25\% of total TAC members (per AO) must be from Russian Federation scientific community.} and the US.  An increasing trend in the fraction of female \ac{TAC} members from $\sim$0.2 in AO-1 to $\sim$0.4 by AO-19 is evident. Given the small number of panel chairs (59 in a total of 19 AOs), it is more difficult to quantify trends. However, it is worth noting that there were no female \ac{TAC} panel chairs until AO-8 in 2010. 

\section{\textbf{INTEGRAL Users Group}}
\label{subsec:IUG}

\begin{figure*}[ht]
\centering
\includegraphics[width=1.0\textwidth,angle=0]{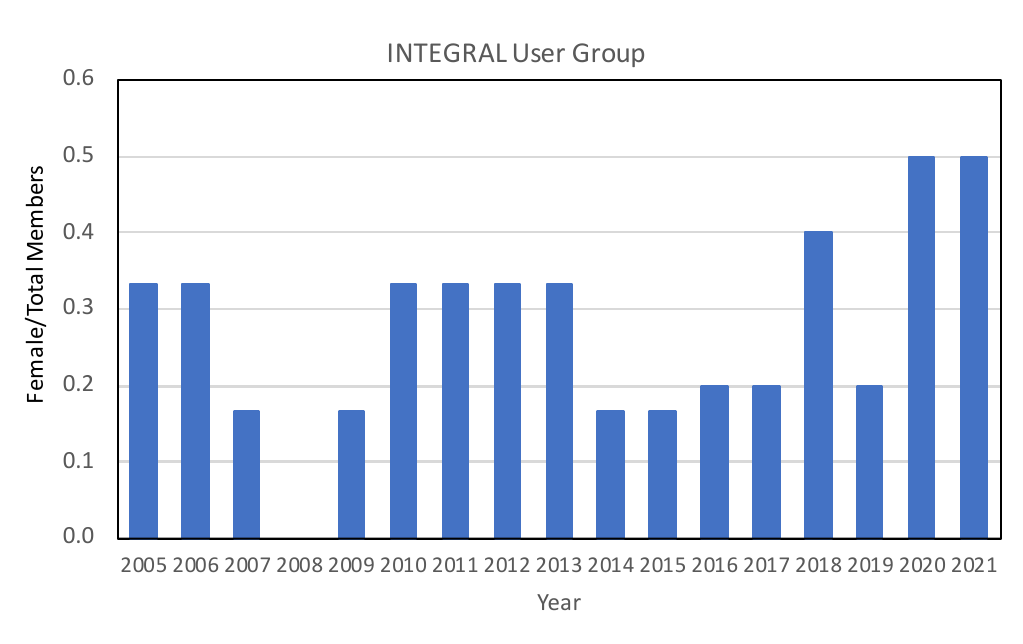}
\caption{The fraction of female INTEGRAL \ac{IUG} members compared to the total membership between 2005 and 2021.}
\label{fig:INTEGRAL UG}    
\end{figure*}

As well as the \ac{TAC}, there is another body involved in optimising the scientific output of the mission. This is the INTEGRAL Users Group (\ac{IUG}) which advises ESA, through the Project Scientist, on all matters relating to the optimisation of the scientific output of the mission. It also acts as a forum to discuss input from the community of users, and when appropriate advise or recommend action to \ac{ESA} regarding INTEGRAL operations. The \ac{IUG} is smaller than the \ac{XUG} with only 6 members (including the chair). The names of the members are made public. The \ac{IUG} was established in 2005, in parallel to the already existing \ac{ISWT}. As a consequence of the termination of the \ac{CP} beyond 2008, the \ac{IUG} and \ac{ISWT} teams were merged in November 2007 into one \ac{IUG}. The chair of the INTEGRAL \ac{TAC} is also a member of the \ac{IUG} to ensure good coordination between the two bodies. In addition, the INTEGRAL Project Scientist, Mission Manager and instrument principal investigators attend the meetings. Note that there was a change in Project Scientist in 2013.
Fig.~\ref{fig:INTEGRAL UG} shows the number of female \ac{IUG} members divided by the total between 2005 and 2021. On average 28\% of the total number of members were female with no obvious trend with time. We note that 53\% of the \ac{IUG} chairs were female.

\section{\textbf{Proposers}}

Between AO-1 and AO-19 there were 2115 proposals submitted in response to INTEGRAL calls for observing opportunities. This provides a rich data set to quantify many aspects of the proposal process. When submitting a proposal, INTEGRAL proposers are required to indicate the country where their institute or university is located. These are selected from a drop-down menu. This information allows an examination of the nationality distribution of the proposers' institutes. Proposers are not required to submit information on gender, age or type of position held and these have to be derived by other means if required.

\subsection{\textbf{Proposers' Institute Countries}}

Table~\ref{tab:INTEGRAL_ProposalNationalities} shows the number of proposals submitted and accepted for a range of countries including all the \ac{ESA} Member States that submitted at least one proposal and other countries from which significant numbers of proposals originated.  Only eight proposals were received from countries that are not listed in Table~\ref{tab:INTEGRAL_ProposalNationalities}.

\begin{table}
\centering
\caption{The number of proposals submitted and accepted from scientists located in different countries.}
\begin{tabular}{lcccc}
\hline \noalign{\smallskip}
\vspace{3pt}
Country	& No. Proposals	& Percentage	&No. &	Acceptance \\
&	Submitted&	of Total&	Accepted&	Percentage \\
\hline \noalign{\smallskip}
BELGIUM	& 8 &	0.38 &	2	& 25.0 \\
CZECH REPUBLIC &	3 &	0.14 & 0 &	0.0\\
DENMARK &	35	& 1.65 &	4	& 11.4\\
FINLAND	& 5 &	0.24 &	4 &	80.0\\
FRANCE	& 245 &	11.58 &	94 &	38.4\\
GERMANY	& 328 &	15.51 &	155 &	47.3\\
GREECE	& 1	& 0.05 &	0 &	0.0\\
IRELAND	& 28	& 1.32 & 	1	& 3.6\\
ITALY &	472 &	22.32 &	169 &	35.8\\
NETHERLANDS	& 70 &	3.31 &	33 & 	47.1\\
POLAND	& 18	& 0.85 &	6	& 33.3\\
SPAIN &	142 &	6.71& 	88&	62.0\\
SWEDEN &	4	& 0.19 &	0	& 0.0\\
SWITZERLAND	& 149 &	7.04 &	45 &	30.2\\
UNITED KINGDOM	& 74 &	3.50 &	33 &	44.6\\
AUSTRALIA &	9 &	0.43 &	6 &	66.7\\
CHINA	& 21 &	0.99 &	10 &	47.6\\
INDIA	& 2 &	0.09 &	0 &	0.0\\
ISRAEL	& 1 &	0.05 &	0 &	0.0\\
JAPAN	& 18 &	0.85 &	4 &	22.2\\
MEXICO	& 1	& 0.05 &	0	& 0.0\\
RUSSIAN FEDERATION	& 189 &	8.94 & 101 &	53.4\\
TURKEY	& 14 &	0.66 &	9 &	64.3\\
UNITED STATES& 	270 &	12.77 &	107 &	39.6\\
\hline \noalign{\smallskip}
\end{tabular}
\label{tab:INTEGRAL_ProposalNationalities}
\end{table}

\begin{figure*}[ht]
\centering
\includegraphics[width=1.0\textwidth,angle=0]{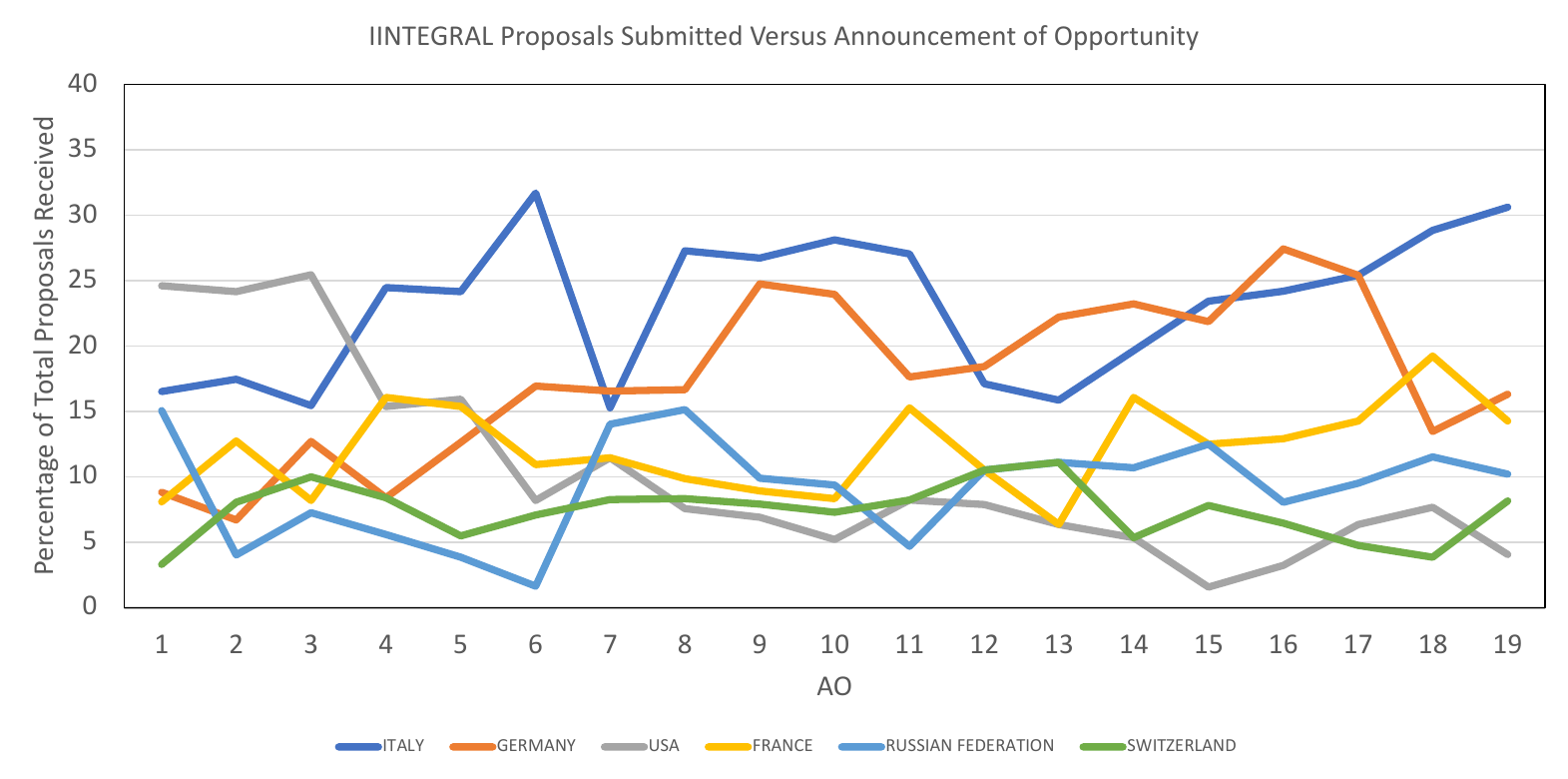}
\caption{Proposals submitted from the six countries with the largest numbers of proposals against AO number. There are no large changes in the percentages during the mission, suggesting that the level of interest from the six countries was unchanged. The only exception is a decline in the percentage of proposals from PIs located in the USA (grey line) as the mission progressed. Note that the Guaranteed Time ended in 2008 around the time of AO-6.}
\label{fig:INTEGRAL_ProposalNationalities}    
\end{figure*}

Scientists located at institutes and universities within Italy submitted the most proposals -- nearly a quarter of the total (22.3\%). This is followed by Germany (15.5\%), the USA (12.8\%), France (11.6\%), the Russian Federation (8.9\%), and Switzerland (7.0\%). Figure~\ref{fig:INTEGRAL_ProposalNationalities} shows the percentages of proposal numbers versus \ac{AO} for the six countries whose scientists submitted the most proposals. These show relatively stable proposal submission fractions with \ac{AO} number. There is evidence for a decline in the percentage of proposals from \ac{PI}s located in the USA as the mission progressed (grey line in Fig.~\ref{fig:INTEGRAL_ProposalNationalities}).  This may indicate a lessening of interest in the mission from US based PIs as the mission progressed. We note that \ac{NASA} funding for US INTEGRAL proposals was only available from AO-1 to AO-7. 

In order to be able to draw reliable conclusions, we examined the number of accepted proposals for the seven countries whose scientists have submitted at least 100 proposals (Table~\ref{tab:INTEGRAL_ProposalNationalities}). the countries with the highest accepted fractions are Spain where 62.0\% of 142 proposals were accepted, followed by the Russian Federation with 53.4\% accepted of 189 proposals, and Germany with 47.3\% accepted of 328 proposals.  

\subsection{\textbf{Proposers' Gender}}

\begin{figure*}[ht]
\centering
\includegraphics[width=1.0\textwidth,angle=0]{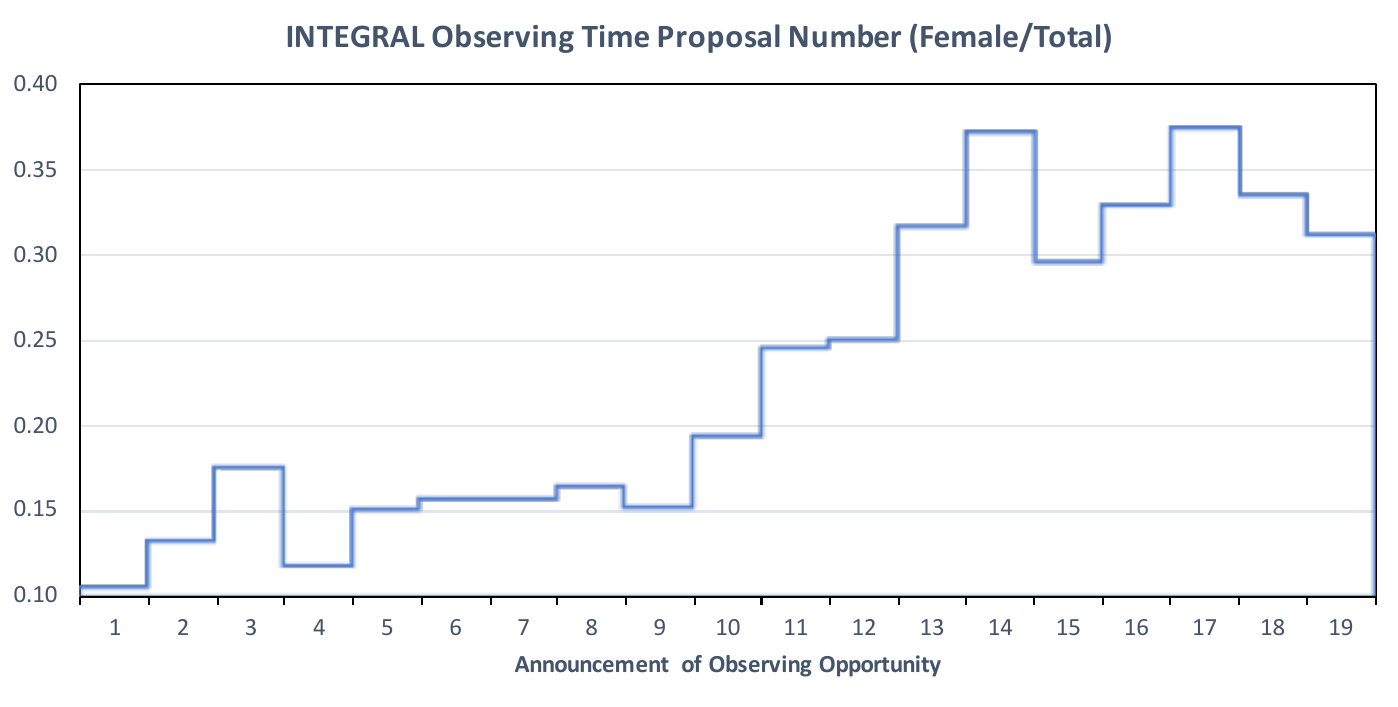}
\caption{The fraction of submitted INTEGRAL proposals with female PIs compared to the total. This increased from $\sim$10\% to $\sim$30\% between AO-1 and AO-19.}
\label{fig:INTEGRAL_FemaleProposers}    
\end{figure*}

Figure~\ref{fig:INTEGRAL_FemaleProposers} shows that the fraction of proposals with female \ac{PI}s compared to the total increased from $\sim$10\% to $\sim$30\% between AO-1 and AO-19.

We next examined the number of proposals with male and female \ac{PI}s from the same seven countries (Table~\ref{tab:INTEGRAL_ProposalNationalities>100}). Since proposers are not asked to specify their gender (nor age or type of position held etc.) gender information for each proposer was obtained through examining publicly-accessible web-based data in a similar way as for the \ac{HST} study by I.N.\ Reid ~\cite{2014Reid} and from knowledge of the community they serve by the Project Scientist and \ac{SOC} staff. We appreciate that gender identity is more complex than a binary issue, however, no attempt can be made to assign genders other than male or female as this information is not readily available. 

\begin{figure*}[ht]
\centering
\includegraphics[width=1.0\textwidth,angle=0]{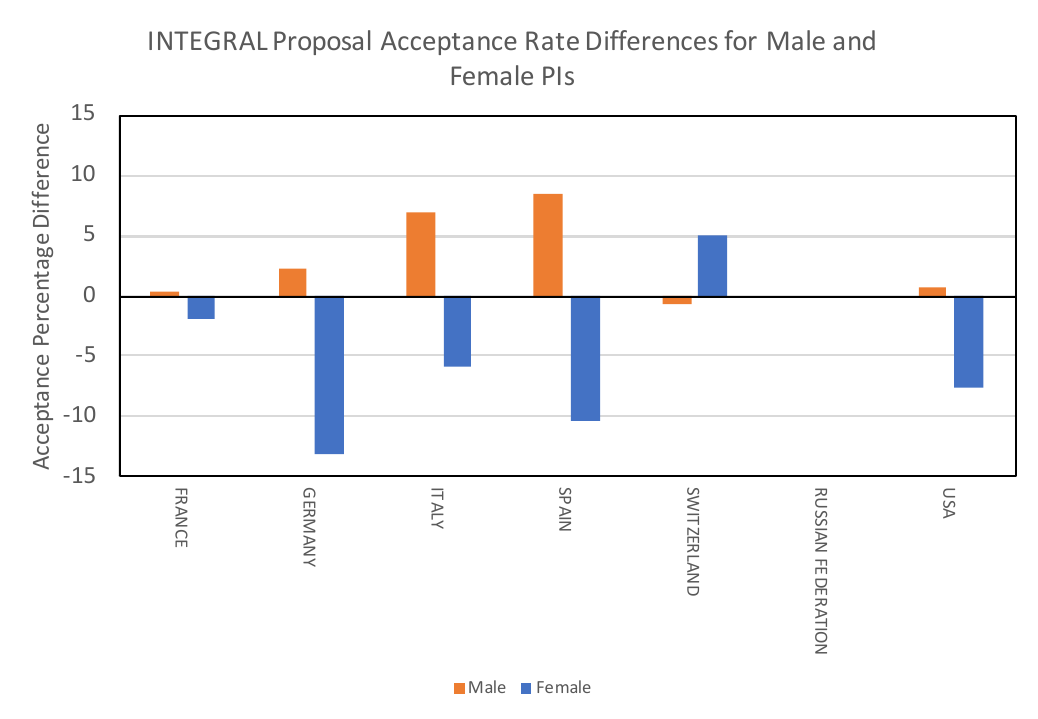}
\caption{The acceptance fraction differences of INTEGRAL proposals with male and female PIs compared to the average for all proposers from that country. The countries are those with at least 100 submitted INTEGRAL proposals.}
\label{fig:INTEGRAL_Proposal_AcceptanceGender}    
\end{figure*}

The country with the highest fraction of female \ac{PI}s is Italy (54.4\%) followed by Spain (45.1\%). The countries with the lowest fraction of female PIs are the Russian Federation (0.0\%) and the USA (9.3\%).  Figure~\ref{fig:INTEGRAL_Proposal_AcceptanceGender} shows the acceptance fraction differences for male and female \ac{PI}s compared to the average for that country for each of the seven countries with $>$100 proposals. It can be seen that Switzerland is the only country to have a better female PI acceptance rates of +5.1\% compared to the average for all Swiss proposals. Female PIs located in Germany and Spain have acceptance rates of $-$13.2\% and $-$10.4\%, respectively compared to the averages for these countries. The accepted fraction difference for the Russian Federation comes out to 0, because the number of female PIs has been 0; see Table~\ref{tab:INTEGRAL_ProposalNationalities>100}.

\begin{table}
\caption{Proposal statistics for PIs located in countries with $>$100 proposals. The \% differences are with respect to the acceptance percentages given in Table~\ref{tab:INTEGRAL_ProposalNationalities} which are for all proposals from each country}
\setlength\tabcolsep{4.5pt}
\begin{tabular}{lccccccc}
\hline \noalign{\smallskip}
\vspace{3pt}
Country	& \multicolumn{2}{c}{Submitted} & Female& \multicolumn{2}{c}{Accepted} & \multicolumn{2}{c}{\% Difference} \\
&	Male PI & Fem. PI & /Total (\%) & Male PI & Fem. PI & Male PI & Fem. PI	 \\
\hline \noalign{\smallskip}
FRANCE & 209 & 36 & 14.7 & 80 & 13 & 0.3 & $-$1.8 \\
GERMANY & 281 &	47 & 14.3 & 139 & 16 & 2.2 & $-$13.2 \\
ITALY & 215 & 257 &	54.4 & 92 & 77 & 7.4 & $-$5.8 \\
SPAIN & 78 & 64 & 45.1 & 55 & 33 & 8.5 & $-$10.4 \\
SWITZERLAND & 132 & 17 & 11.4 & 39 & 6 & $-$0.7 & 5.1 \\
RUSSIAN FEDERATION & 189 & 0 & 0.0 & 101 & 0 & 0.0 & ... \\
USA & 245 & 25 & 9.3 & 98 & 8 & 0.4 & $-$7.6 \\
\hline \noalign{\smallskip}
\end{tabular}
\label{tab:INTEGRAL_ProposalNationalities>100}
\end{table}

\section{\textbf{Proposal Selection}}

We decided to only investigate the outcomes of proposals that requested observing time -- so data rights or similar proposals were not included in the analysis of success rates. 
Data rights proposals have different factors affecting their selection and so their success rates cannot be directly compared with other proposals. In addition, data right proposals are a peculiarity of the INTEGRAL mission and cannot easily be compared with other missions.  We note that, examination of the 607 data rights proposals shows that 511 were awarded data rights. This is 84\% of the proposals, a much higher percentage than for proposals requesting dedicated observing time. For the proposals requesting observing time, we investigated the number of accepted proposals and the amount of observing time allocated.  We examined the statistics for all proposals -- Priority A, B and C and only for those ranked in Priority A and B by the \ac{TAC} since these are of major scientific importance and had a much higher chance of being scheduled than the "filler" Priority C targets. 

\subsection{\textbf{Proposers' Age}}
 
In order to determine the "academic age" of proposers we used the difference between the year that their \ac{PhD} was awarded and the year that an AO to which they submitted a proposal was issued. Thus a proposer who submits multiple proposals to the same AO will be counted multiple times in the same "academic year", whilst one who submits multiple proposals to different AOs will be counted in different "academic years". 

The year the proposer obtained their \ac{PhD} (or equivalent) were determined for 99.7\% of the proposals by searching the internet, particularly sites such as the \ac{ADS}, LinkedIn, the Astronomy Genealogy Project (\url{astrogen.aas.org}), \url{ORCID.org}, IEEE Xplore (\url{https://ieeexplore.ieee.org/Xplore/home.jsp}), and, for French theses \url{https://www.theses.fr}. Some proposers were contacted and provided their \ac{PhD} dates by email. The proposers for which the \ac{PhD} could not be found are often retired, deceased or have left astronomy for some other reason. For \ac{PhD} students who had not yet completed their degrees, the expected year of submission was used. For the small number of proposers who did not have a PhD and were not enrolled in a \ac{PhD} programme, their dates were assumed to be arbitrarily far in the future. For some senior scientists in Italian institutes who do not have a \ac{PhD}, three years after they obtained their Laurea was used. It should be noted that using the year of \ac{PhD} to indicate the number of years experience neglects time spent outside of astronomy such as as a carer. We note that the "youngest" proposer was 13 years before obtaining a \ac{PhD} and the oldest 53 years after.

\begin{table}
\centering
\caption{INTEGRAL AO proposal numbers showing for proposals with male and female PIs and the number of proposals accepted. Only proposals that requested observing time are included.}
\begin{tabular}{lccccc}
\hline \noalign{\smallskip}
\vspace{3pt}
AO & \multicolumn{3}{c}{Submitted} & \multicolumn{2}{c}{Accepted} \\
 & No. of & Male PI & Female PI &  Male PI & Female PI \\
 & Proposals  & Proposals  & Proposals  & Proposals & Proposals \\
\hline \noalign{\smallskip}
1 & 265 & 236 & 29 & 75 & 15 \\
2 & 143 & 124 & 19 & 71 & 12 \\
3 & 103 & 84 & 19 & 48 & 11 \\
4 & 93 & 82 & 11 & 40 & 5 \\
5 & 93 & 79 & 14 & 49 & 5 \\
6 & 70 & 59 & 11 & 35 & 8\\
7 & 70 & 59 & 11 & 38 & 5 \\
8 & 61 & 51 & 10 & 32 & 6 \\
9 & 46 & 39 & 7 & 33 & 5 \\
10 & 57 & 46 & 11 & 31 & 7 \\
11 & 53 & 40 & 13 & 32 & 10 \\
12 & 72 & 54 & 18 & 29 & 8\\
13 & 57 & 39 & 18 & 23 & 9 \\
14 & 54 & 34 & 20 & 27 & 11\\
15 & 61 & 43 & 18 & 31 & 9\\
16 & 58 & 39 & 19 & 27 & 7 \\
17 & 59 & 37 & 22 & 29 &14 \\
18 & 48 & 32 & 16 &27 & 10 \\
19 & 45 & 31 & 14 & 27 & 9\\
\textbf{Total} &\textbf{1508} & \textbf{1208} & \textbf{300} & \textbf{704} & \textbf{166} \\
\hline \noalign{\smallskip}
\end{tabular}
\label{tab:INTEGRAL_AONumbers}
\end{table}

\subsection{\textbf{Priority A, B and C Proposal Acceptance Rates}}

Of the 1508 proposals which requested observing time that were submitted to AO-1 to AO-19, 870 were awarded an amount of observing time. This corresponds to an overall success rate of 57.7\%.  This is higher than for XMM-Newton (see~\cite{2024XMM}), consistent with the more moderate over-subscription factors given in Table~\ref{tab:INTEGRAL_AO} compared to those in~\cite{2024XMM}.  We next examined the number of proposals for each gender that were awarded any observing time. A total of 1210 such proposals with male \ac{PI}s and 298 proposals with female \ac{PI}s were submitted, of which 704 and 166 were awarded observing time. This gives success rates of 58.2\% and 55.7\% for male and female PIs, respectively. This is a difference in favour of males of 4.4\%. 

Proposal submissions and proposal acceptance are unlikely to be independent or random processes and for Poisson statistics to apply events need to be independent of each other.
If this is assumed to be the case then the success rates are $58.2 \pm 2.8$\% and $55.7 \pm 5.4$\% for male and female \ac{PI} proposals, respectively and a difference in favour of males of $4.4 \pm 6.7$\% corresponding to a male to female success ratio of $1.053 \pm 0.113$. These uncertainties are given only to illustrate the outcomes if Poisson statistics were to apply to the selection process. We note that actual uncertainties on the proposal numbers for each AO are zero. We further note that there are systematic uncertainties associated with this process due to misappropriated genders and incorrect \ac{PhD} "academic ages", but these are likely to be too small to significantly affect our calculations.

\begin{figure*}[ht]
\centering
\includegraphics[width=1.0\textwidth,angle=0]{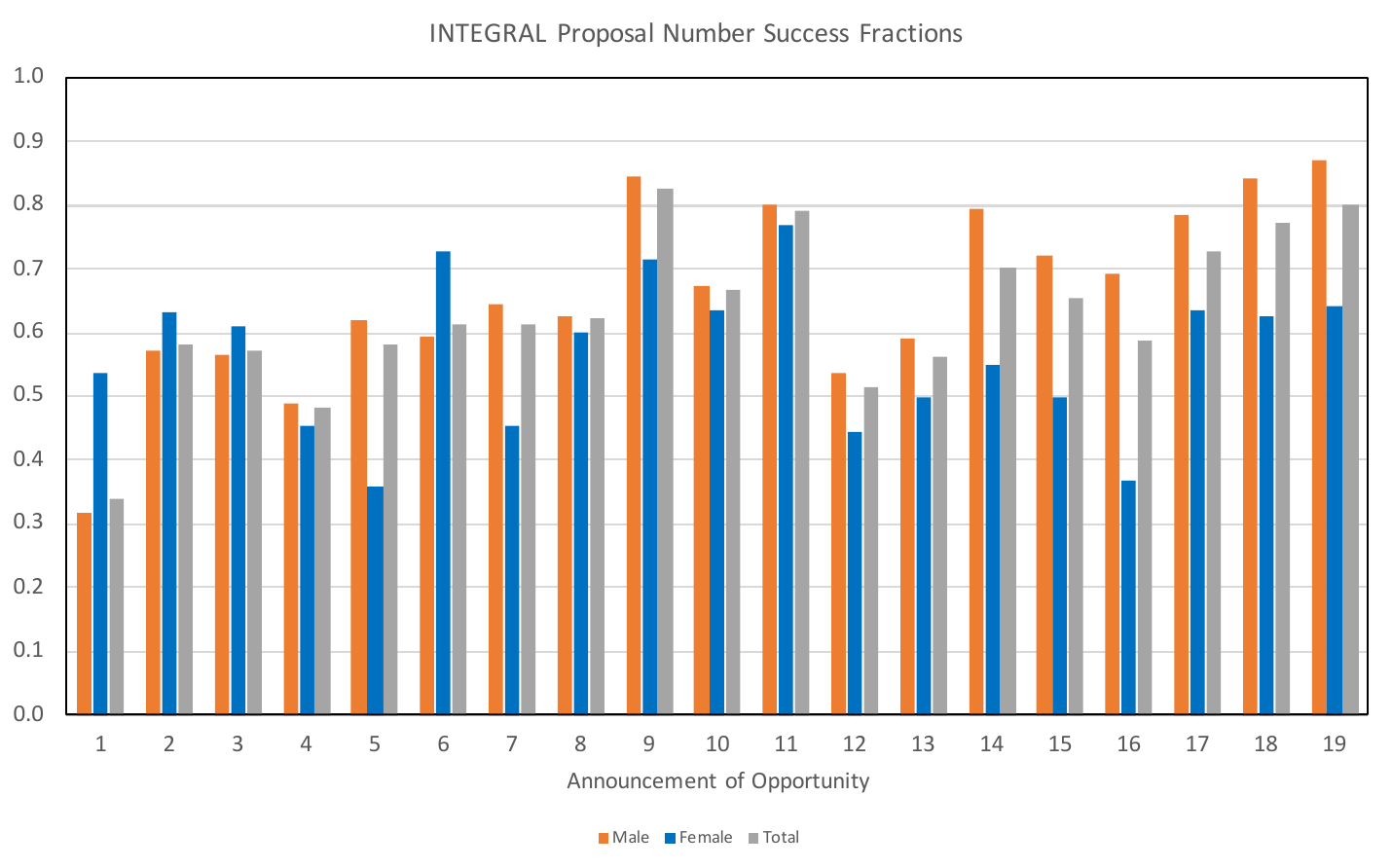}
\caption{The fraction of accepted proposals compared to the number submitted. The histograms show the acceptance fractions for male \ac{PI}s, female \ac{PI}s and all \ac{PI}s.}
\label{fig:INTEGRAL_ProposalNumbers}    
\end{figure*}

The success rates are shown for male, female and all proposers for each \ac{AO} in Fig.~\ref{fig:INTEGRAL_ProposalNumbers}. In contrast to the results from \ac{HST} Cycles 11 to 20 reported in \cite{2014Reid} where proposals with male \ac{PI}s had a consistently higher success rate than those with female \ac{PI}s, no such trend is visible in INTEGRAL proposals. There is no obvious evolution in the male and female PI proposal acceptance rates with \ac{AO} number, which cover an interval of more than 20 years. 

\begin{figure*}[ht]
\centering
\includegraphics[width=1.0\textwidth,angle=0]{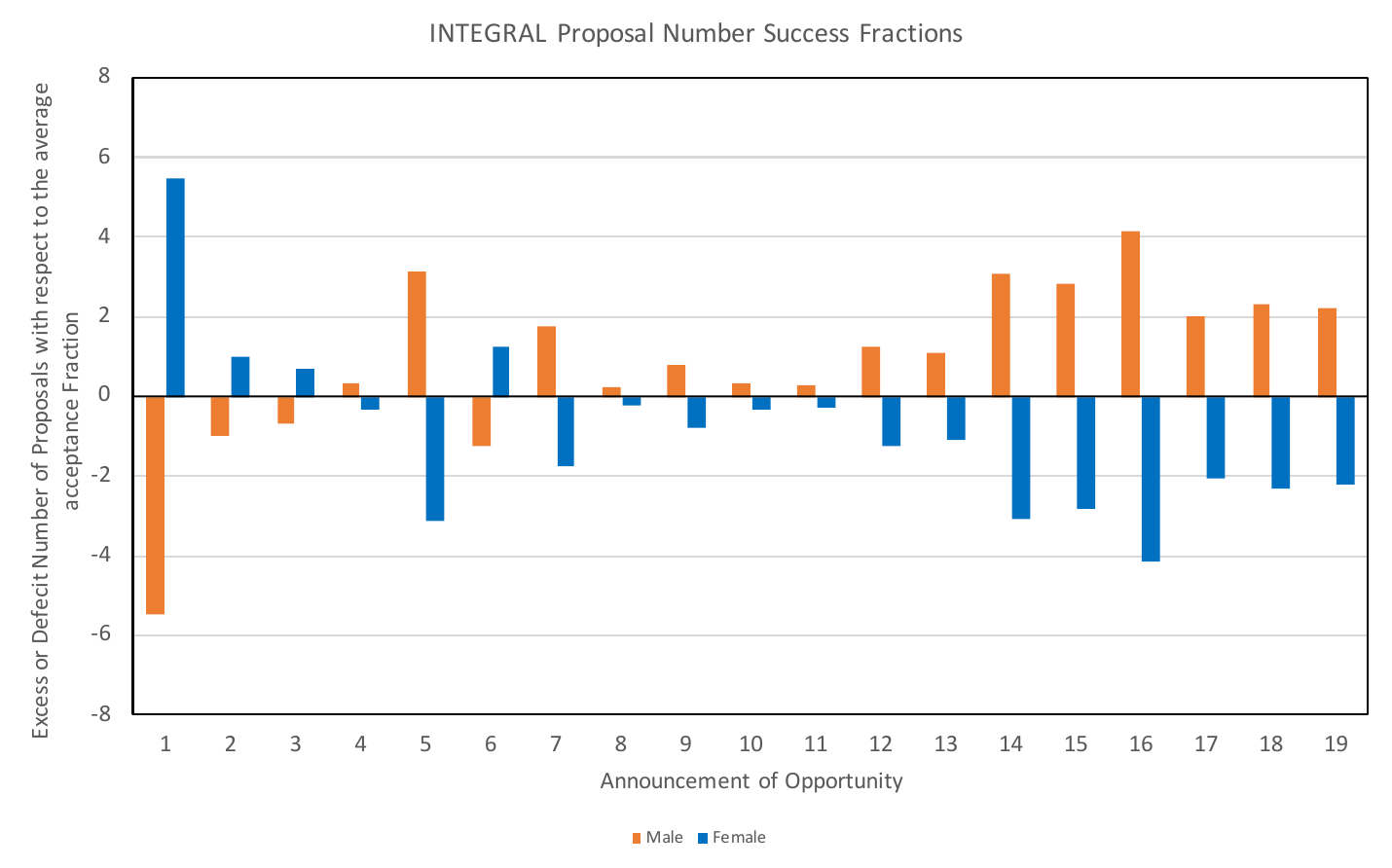}
\caption{The relative success rates of INTEGRAL proposals.The histograms show the difference between the actual number of successful proposals and the expected number based on the overall acceptance rate.}
\label{fig:INTEGRAL_Proposal_Excess}    
\end{figure*}

Figure~\ref{fig:INTEGRAL_Proposal_Excess} shows the variation in acceptance rates more clearly. For each \ac{AO}, it shows the difference between the expected number of accepted proposals, calculated using the overall acceptance rate, and the number actually accepted. It shows that the largest discrepancies occurred during AO-1 and AO-2 where proposals with female \ac{PI}s had higher success rates of 17.9\% and 6.6\%, respectively, compared to proposals with a male \ac{PI}. The most successful \ac{AO} for male \ac{PI} proposals was AO-19 where proposals led by male \ac{PI}s were 4.9\% more likely to be accepted than those with a female \ac{PI}. 

\begin{figure*}[ht]
\centering
\includegraphics[width=1.0\textwidth,angle=0]{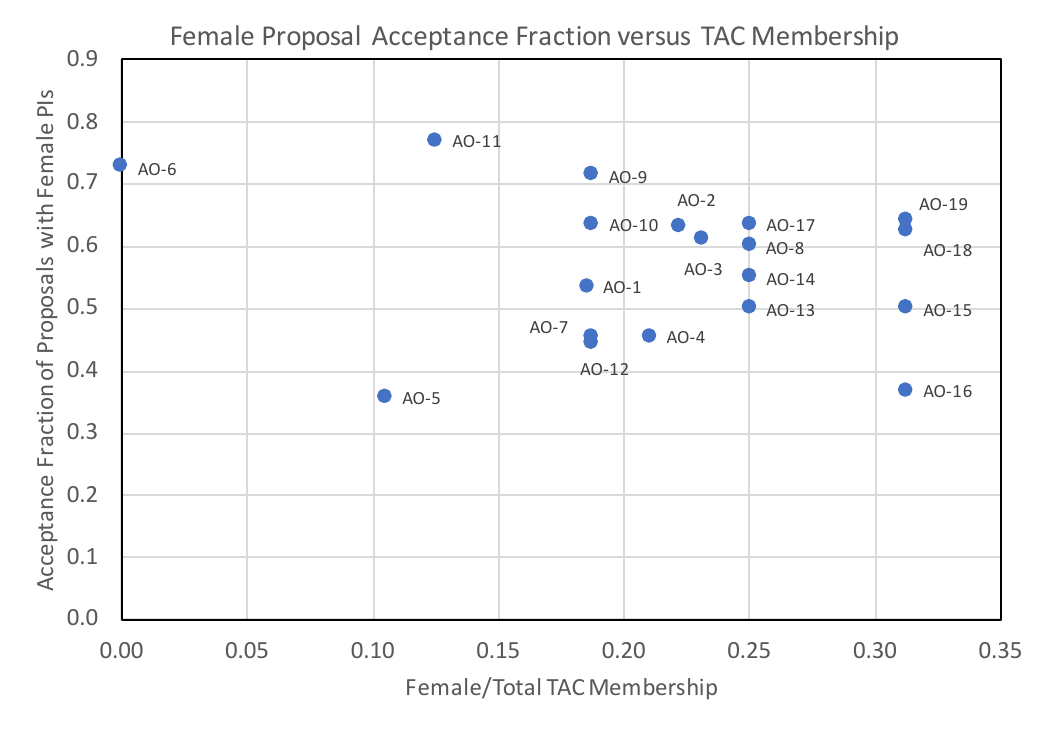}
\caption{The acceptance fraction for female PIs for each AO (labelled) against the fraction compared to the total of female TAC members. There is no obvious correlation between the two, or with the number of female panel chairs.}
\label{fig:INTEGRAL Acc v TAC}    
\end{figure*}

We examined the \ac{TAC} membership genders to see if these are correlated with proposal acceptance rates. Figure~\ref{fig:INTEGRAL Acc v TAC} shows the acceptance fraction of proposals with female \ac{PI}s against the fraction of female compared to the total number of \ac{TAC} members. It can be seen that there is no obvious correlation between the success rates of female-led proposals and the fraction of female \ac{TAC} members.
We note that there were no female chairs for AO-1 to AO-8, or AO-13 and AO-14 (Fig.~\ref{fig:INTEGRAL TAC Chairs}) and the results from these \ac{AO}s are indistinguishable from the other \ac{AO}s suggesting that having females in leadership positions in the \ac{TAC} does not make a measurable difference to female PI proposal success rates. 

\begin{figure*}[ht]
\centering
\includegraphics[width=1.0\textwidth,angle=0]{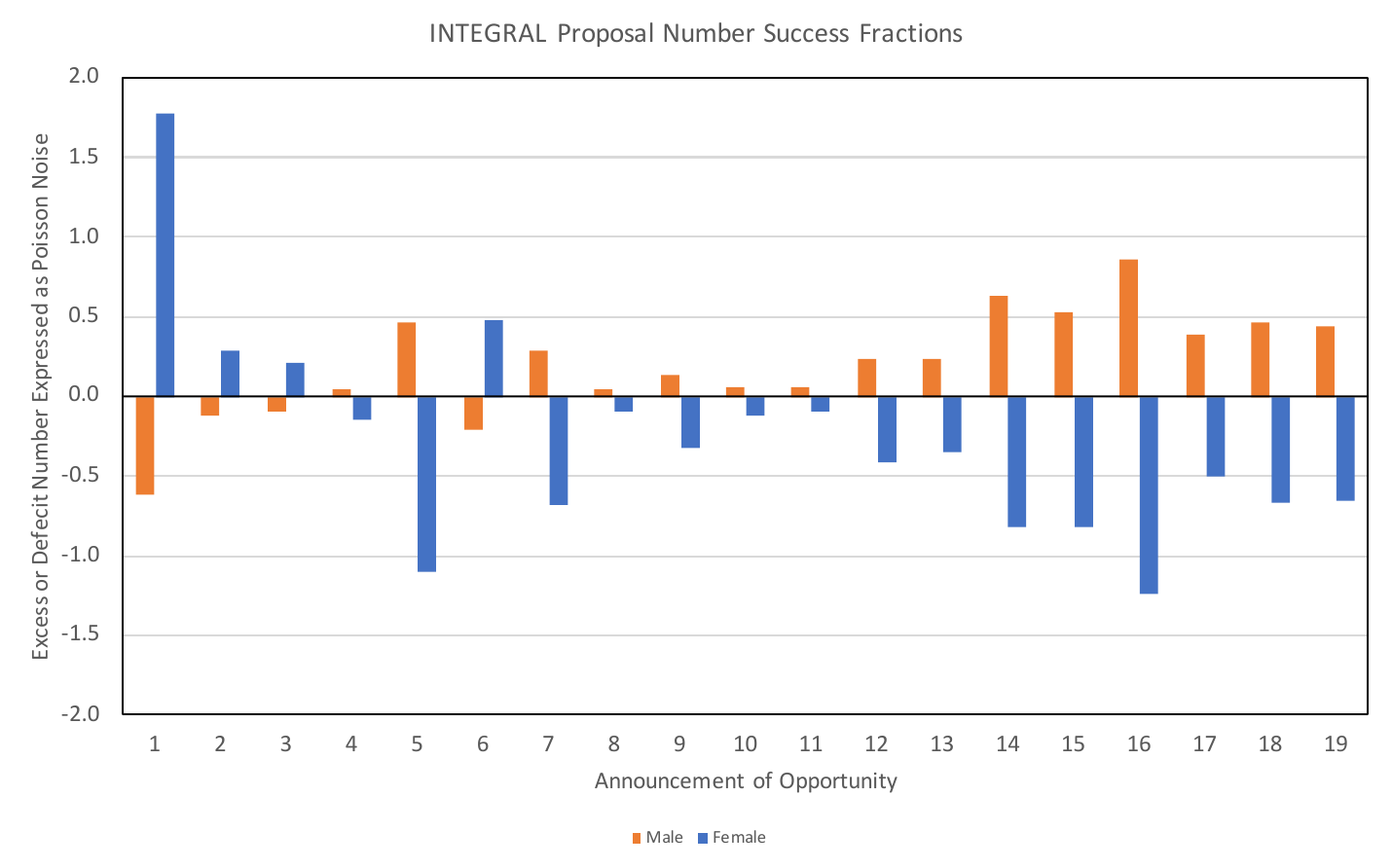}
\caption{The relative success rates of INTEGRAL proposals. The histograms show the difference between the actual number of successful proposals and the expected number based on the overall acceptance rate normalised by dividing by the square root of the number of expected proposals.}
\label{fig:INTEGRAL_Proposal_Poisson}    
\end{figure*}

Figure~\ref{fig:INTEGRAL_Proposal_Poisson} shows the same excesses depicted as a quasi-Poisson quantity, normalised by dividing by the square root of the expected number of accepted proposals. While Poisson statistics do not necessarily apply to the peer-review process, scientists often adopt such a statistic in assessing the difference between observation and prediction. It is interesting to note that from the 19 \ac{AO}s examined, nine had higher female \ac{PI} success rates than for male \ac{PI}s, while the opposite occurred during 10 \ac{AO}s.

\begin{table}
\centering
\caption{INTEGRAL requested and accepted observing time for male and female proposers between AO-1 and AO-19. The last two columns give the ratio of accepted to proposed observing time for each gender. A proposal with a male \ac{PI} which requested 1000~Msec of observing time in AO-1 has been excluded.}
\begin{tabular}{lcccccc}
\hline \noalign{\smallskip}
\vspace{3pt}
AO & \multicolumn{4}{c}{Time (Msec)} & \multicolumn{2}{c}{Success Rate}\\
      &  \multicolumn{2}{c}{Requested} & \multicolumn{2}{c}{Accepted} & \multicolumn{2}{c}{Time} \\
      &  Male & Female & Male & Female & Male & Female \\
\hline \noalign{\smallskip}
1 & 283.0 &39.8&	32.7&	5.4&	0.116&	0.135 \\
2 & 132.4&	14.1&	49.4&	5.7&	0.373&	0.402  \\
3 & 102.3&	12.7&	48.3&	6.7&    0.472&	0.526   \\
4 & 121.0&	15.5&	27.6&	6.3&	0.228&	0.406  \\
5 & 132.9&	23.6&	46.0&	6.4&	0.346&	0.272  \\
6 & 75.0&	11.6&	26.8&	5.1& 	0.357&	0.439  \\
7 &	92.5&	11.3&	41.0&	5.7&	0.443&	0.499 \\
8 & 71.9&	14.1&	33.1&	7.4&	0.461&	0.527  \\
9 & 52.3&	10.1&	32.9&	7.1&	0.628&	0.701   \\
10 & 48.7&	9.8&	30.2&	6.8&	0.619&	0.698  \\
11 & 36.1&	14.5&	29.0&	12.2&	0.805&	0.838  \\
12 & 77.6&	21.6&	33.3&	7.4&	0.430&	0.342  \\
13 & 81.0&	21.7&	27.7&	9.0&	0.342&	0.415  \\
14 & 74.3&	37.2&	25.6&	9.6&	0.345&	0.258 \\
15 & 47.9&	33.5&	34.8&	8.6&	0.726&	0.258 \\
16 & 76.3&	33.2&	32.1&	7.9&	0.421&	0.237 \\
17 & 44.5&	36.0&	32.7&	12.0&	0.735&	0.332 \\
18 & 38.2&	36.2&	35.3&	10.5&	0.923&	0.290\\
19 & 38.3&	26.8&	37.2&	13.2&	0.971&	0.492\\
\textbf{Total} &\textbf{1626.2} & \textbf{423.2} & \textbf{655.6} & \textbf{152.7} &  \\
\hline \noalign{\smallskip}
\end{tabular}
\label{tab:INTEGRAL_AOResults}
\end{table}

\subsection{\textbf{Priority A and B Proposal Acceptance Rates}}

Of the 1508 proposals which requested observing time that were submitted to AO-1 to AO-19, 690 were awarded Priority A or B observing time. This corresponds to an overall success rate of 45.8\%. We next examined the number of proposals for each gender that were awarded high-priority observing time. A total of 1210 such proposals with male \ac{PI}s and 298 proposals with female \ac{PI}s were submitted, of which 556 and 134 proposals were awarded observing time. This gives success rates of 46.0\% and 45.0\% for male and female PIs, respectively. This is a difference in favour of males of 2.2\%. Again, assuming square root errors on the numbers of submitted and accepted proposals gives success rates of $46.0 \pm 2.4$\% and $45.0 \pm 4.7$\% for male and female \ac{PI} proposals, respectively. This corresponds to a male to female success ratio of $1.022 \pm 0.118$.

\subsection{\textbf{Priority A, B and C Observing Time Acceptance Rates}}
\label{subsec:INTEGRALRates}

We examined the acceptance rates for the genders using awarded times. Many INTEGRAL proposals are allocated less observing time than requested, often because not all the requested targets are approved. If the allocation of observing time is handled differently for males and female \ac{PI}s, then this will show as a difference in the relative amounts of time approved, compared to the number of proposals accepted. The results for each \ac{AO} are summarised in Table~\ref{tab:INTEGRAL_AOResults}.  This shows the total time requested by proposals with male and female \ac{PI}s to be 2626.2 and 423.2~Msec, respectively. 655.6~Msec and 152.7~Msec of observing time was awarded to proposals with male and female \ac{PI}s, respectively. This gives average success rate of 40.3\% for male \ac{PI}s and 36.1\% for female \ac{PI}s. This is a difference of 11.5\% in favour of male \ac{PI}s. We note that a single proposal with a male \ac{PI} which requested 1000~Msec of observing time (or about 32 years!) and was awarded 1~Msec (in Priority C in AO-1) has been excluded from this analysis as the request was deemed unrealistic. 

\begin{figure*}[ht]
\centering
\includegraphics[width=1.0\textwidth,angle=0]{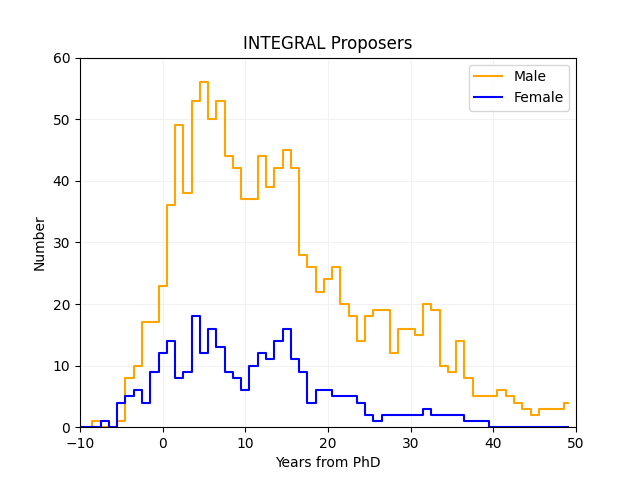}
\caption{The "academic age" distribution (Years since PhD) for male and female INTEGRAL \ac{PI}s. \ac{PI}s have a mean "academic age" of 13.7 years. Male PIs have a mean "academic age" of 14.7 years compared to 10.8 years for female \ac{PI}s.}
\label{fig:INTEGRAL_ProposalAge}    
\end{figure*}

\begin{figure*}[ht]
\centering
\includegraphics[width=1.0\textwidth,angle=0]{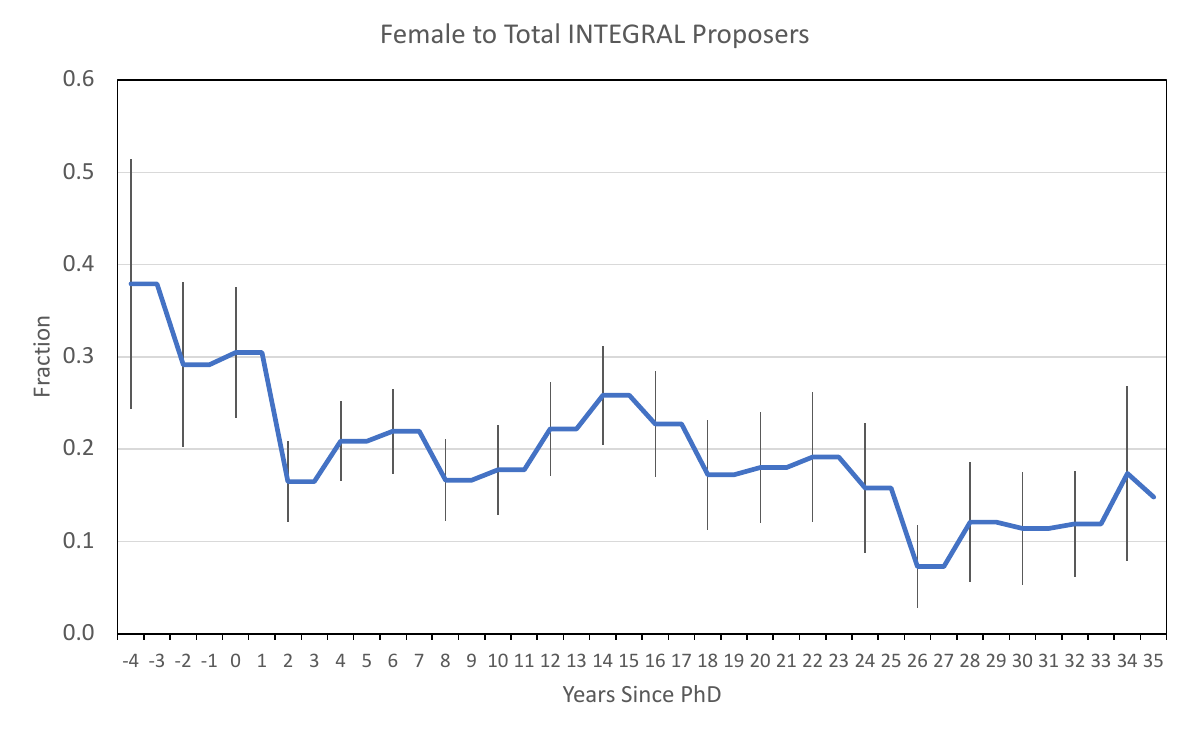}
\caption{The ratio of the "academic age" (years from \ac{PhD}) of female to all INTEGRAL \ac{PI}s in two year bins showing an increase in early career (less than around 5 years after obtaining a \ac{PhD}) female \ac{PI}s compared to the total. The indicative error bars show 1$\sigma$ standard deviations assuming that the number of proposers in each age bin follows a Poisson distribution.
}
\label{fig:INTEGRAL_ProposalAgeRatio}    
\end{figure*}

Figure~\ref{fig:INTEGRAL_ProposalAge} shows the distributions of male and female INTEGRAL \ac{PI}s who obtained their \ac{PhD}s or equivalent in one year bins between 10 years before the PhD date to 50 years after. The male \ac{PI}s who requested observing time (ie. data time proposals are excluded) have an mean "academic age" of 14.7 years post-PhD compared to that of the female \ac{PI}s of 10.8 years. The mean for all \ac{PI}s is 13.7 years. A two-sample Anderson-Darling Test (e.g., \cite{2006Babu}) shows that the hypothesis that both samples come from the same underlying population can be rejected at $>$99\% confidence. This difference can be more clearly seen in Fig.~\ref{fig:INTEGRAL_ProposalAgeRatio} which shows the ratio of the two curves. There appears to be an increase in the number of early career (less than around 5 years after obtaining a PhD) female \ac{PI}s compared to males. Beyond this the ratio of females to the total is $\sim$20\%.

\begin{figure*}[ht]
\centering
\includegraphics[width=1.0\textwidth,angle=0]{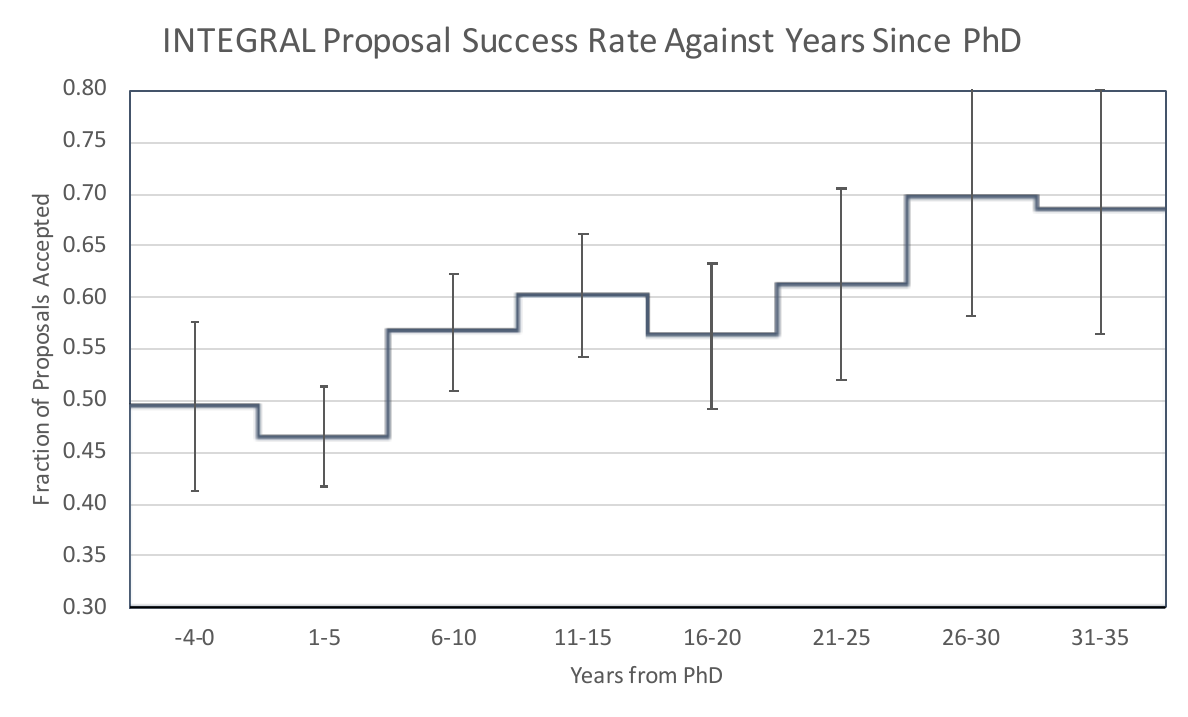}
\caption{The acceptance fraction of all INTEGRAL \ac{PI}s with year of their PhD. This shows an increase in the acceptance rate from $\sim$45\% for \ac{PhD} students to $\sim$70\% for "senior" researchers. The indicative error bars show 1$\sigma$ standard deviations assuming that the number of proposers in each age bin follows a Poisson distribution. 
}
\label{fig:INTEGRAL_ProposalPhD}    
\end{figure*}

We next investigated the acceptance rates for the proposals against PhD year. This is shown for years $-$4 to 35~years after \ac{PhD} in Fig.~\ref{fig:INTEGRAL_ProposalPhD}. A least squares fit to the data gives an intercept of ($45.7 \pm 0.4$)\% and a gradient of ($0.73 \pm 0.26$)\% year$^{-1}$ assuming that the uncertainties are the square root of the number of people in each bin. The value of R$^2$ is 0.45. Indicative error bars show 1$\sigma$ standard deviations assuming that the number of proposers in each age bin follows a Poisson distribution. There as an increase in the acceptance rate from $\sim$45\% for \ac{PhD} students to $\sim$70\% for late career researchers. We note that this gradient is almost a factor two stronger than measured with XMM-Newton. This increase could result from a combination of factors including:

\begin{enumerate}
    \item An increase in the success rate of researchers as they gain experience and have expanded networks of collaborators.
    \item Less successful proposers who leave astronomy.
    \item Less successful proposers who remain in astronomy, but do not propose in subsequent INTEGRAL \ac{AO}s.
    \item A bias in the selection process towards late career scientists.
\end{enumerate}

The positive gradient of this relation indicates that the female early career proposer population is less likely to have proposals accepted simply due to having less experience. To calculate the size of this effect, we took the male and female \ac{PI} distributions shown in Fig.~\ref{fig:INTEGRAL_ProposalAge} and multiplied the number of \ac{PI}s each year by the linear value derived from the fit to the overall success rates. This gave a predicted difference in the acceptance ratio of 2.4\% for male and female \ac{PI}s. This indicates that the underlying "academic age" differences between the two populations may be responsible for about half of the difference in gender acceptance rates of about 4.4\% in proposal numbers or a quarter of the difference in accepted time (11.5\%).   

\begin{figure*}[ht]
\centering
\includegraphics[width=1.0\textwidth,angle=0]{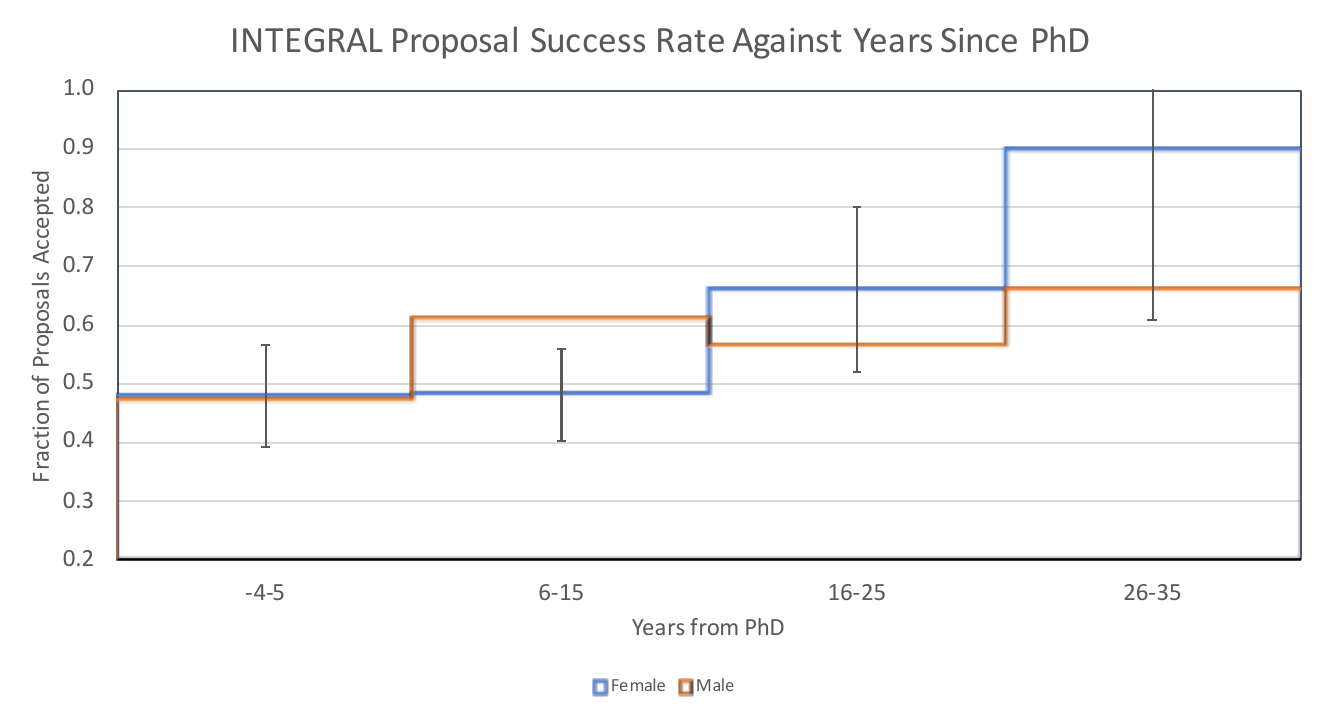}
\caption{The acceptance fractions of female and male INTEGRAL PIs with year of their PhD. The proposal success rate of females PIs tends to increase with "academic age" whilst that of male PIs shows a smaller increase, or may be constant. The indicative error bars show 1$\sigma$ standard deviations assuming that the number of proposers in each age bin follows a Poisson distribution. 
Error bars are only shown for female PIs as these are much larger than for male PIs due to the smaller number of female PIs.
}
\label{fig:INTEGRAL_ProposalPhD_Gender}    
\end{figure*}

The acceptance fractions of female and male INTEGRAL PIs with the binned year of their \ac{PhD} are shown in Fig.~\ref{fig:INTEGRAL_ProposalPhD_Gender}. The proposal success rate of females PIs tends to increase with "academic age" whilst that of male PIs shows a smaller increase, or may be constant. This is in contrast to results reported from \ac{HST} \cite{2014Reid} and XMM-Newton \cite{2024XMM} where the success rates for late career ($\sim$20 years post \ac{PhD}) female PIs appear to be lower than their early career counterparts. Indicative error bars show 1$\sigma$ standard deviations assuming that the number of proposers in each age bin follows a Poisson distribution. Error bars are only shown for female PIs as these are much larger than for male PIs due to the smaller number of female PIs.

\subsection{\textbf{Priority A and B Observing Time Acceptance Rates}}

We next examined the acceptance rates for the genders awarded high-priority (A or B) observing time. The total time requested by proposals with male and female \ac{PI}s is 2626.2 and 423.2~Msec, respectively. 551.1~Msec and 139.2~Msec of Priority A or B observing time was awarded to proposals with male and female \ac{PI}s, respectively. This gives average success rate of 33.9\% for male \ac{PI}s and 32.9\% for female \ac{PI}s. This is a difference of 2.9\% in favour of male \ac{PI}s. As before, a single proposal with a male \ac{PI} which requested 1000~Msec of observing time (or about 32 years!) and was awarded 1~Msec (in Priority C in AO-1) has been excluded from this analysis. 

\section{\textbf{Proposal Selection Summary}}

We have examined the outcomes of 1508 proposals submitted to ESA requesting observing time in response to the AO-1 to AO-19 calls for INTEGRAL observing time. The proposal acceptance rates are summarised in Table~\ref{tab:INTEGRAL_AOSummary}. A total of 58.2\% of proposals with male PIs and 55.7\% with female PIs were awarded {\it any} observing time on INTEGRAL. This is a difference in favour of males of 4.4\%. When the amount of awarded time is evaluated this changes to being 11.5\% in favour of proposals with male \ac{PI}s. This difference implies that proposals with female \ac{PI}s were awarded a smaller fraction of the requested observing time than their male counterparts. For the proposals that were awarded Priority A or B time, the difference in the number of accepted proposals is 2.2\% and in terms of awarded time 2.9\% in favour of male \ac{PI}s. Both of these differences are smaller than when Priority C time is included. This implies that female \ac{PI}s have much lower success rates than their male counterparts for proposals awarded Priority C time. We interpret these four different numbers to imply an overall male to female success rate difference of 2--11\% for INTEGRAL proposals.

The acceptance rate for data rights proposals was 84\% -- much higher than for proposals requesting observing time. This suggests that submitting data rights proposals is a good strategy for obtaining INTEGRAL data. 

Scientists located at institutes within Italy submitted the most proposals -- almost a quarter of the total followed by those located in institutes in Germany, the USA and Switzerland. Amongst the seven countries submitting $>$100 proposals, Spain had 62\% of 142 proposals accepted, followed by the Russian Federation with 53\% of 189 proposals, and Germany with 47\% of 328 proposals. Italy is the country with the highest fraction of proposals from female \ac{PI}s (55\%) followed by Spain (45\%). The countries with the lowest fraction of proposals from female \ac{PI}s are the Russian Federation (0\%) and the USA (9\%). Switzerland is the only country to have a better female \ac{PI} proposal acceptance rate than for male \ac{PI}s of +5\% compared to the average for all Swiss proposals. Proposals from female \ac{PI}s located in Germany and Spain have acceptance rates of $-$13\% and $-$10\%, respectively compared to the averages for their countries.

The fraction of proposals from female \ac{PI}s increased from $\sim$10\% of the total during the early AOs to $\sim$30\% of the total in the later \ac{AO}s. However, there is no marked evolution in the gender balance of the proposal acceptance rates with \ac{AO} number and hence between 2000 and 2021. There was an increase in the fraction of female \ac{TAC} members from $\sim$0.2 of the total for the first \ac{AO}s to $\sim$0.3 for the latest \ac{AO}s and a larger increase in the number of female \ac{TAC} panel chairs from none before AO-9 to one or two out of three chairs in subsequent \ac{AO}s. These results suggest that increasing fractions of female TAC members or panel chairs does not make a measurable difference to the success rate of female PI proposals.

Using the year of obtaining a \ac{PhD}, or equivalent degree, as a proxy for "academic age" shows that the mean "academic age" of female \ac{PI}s is 3.9 years less than for males. A linear fit to the proposal acceptance rate against \ac{PhD} year shows an increase in acceptance rate from $\sim$45\% for students to $\sim$70\% for senior astronomers. The proposal success rate of females \ac{PI}s tends to increase with "academic age" whilst that of male PIs shows a smaller increase, or may be constant. This is in contrast to results reported from \ac{HST} \cite{2014Reid} and XMM-Newton \cite{2024XMM} where the success rates for late career ($\sim$20 years post \ac{PhD}) female PIs appear to be lower than their early career counterparts. 

\begin{table}
\centering
\caption{Summary of INTEGRAL Proposal Acceptance Rates for AO-1 to AO-19 for proposals requesting observing time. A proposal with a male PI which requested 1000~Mec of observing time in AO-1 has been excluded from the analysis. The uncertainties given in brackets assume that the number of submitted and accepted proposals have square root uncertainties and for guidance only.}
\begin{tabular}{llccc}
\hline \noalign{\smallskip}
\vspace{3pt}
Priority &Parameter & \multicolumn{3}{c}{Proposal PI} \\
 && All & Male & Female \\
\hline \noalign{\smallskip}
&Proposals Submitted & 1508 & 1208 & 300 \\
\hline \noalign{\smallskip}
A, B, C &Number Accepted & 870 & 704 & 166 \\
&Percentage Accepted & 57.7  & 58.2  & 55.7  \\
&Ratio Male/Female & & \multicolumn{2}{c}{1.044} \\
A, B & Number Accepted & 690 & 556 & 134 \\
     & Percentage Accepted & 45.8  & 46.0  & 45.0 \\
&Ratio Male/Female & & \multicolumn{2}{c}{1.022} \\
\hline \noalign{\smallskip}
&Time Requested (Msec) &  2049.4 & 1626.2 & 422.7 \\
\hline \noalign{\smallskip}
A, B, C &Time Accepted (Msec) & 808.3 & 655.6 & 152.7 \\
&Percentage Time Accepted & 39.4 & 40.3 & 36.1\\
& Ratio Male/Female & & \multicolumn{2}{c}{1.115} \\
A, B & Time Accepted (Msec) & 690.2 & 551.1 & 139.2 \\
     & Percentage Time Accepted & 33.7 & 33.9 & 32.9 \\
& Ratio Male/Female & & \multicolumn{2}{c}{1.029} \\    
\hline \noalign{\smallskip}
& Overall Male/Female Difference & & \multicolumn{2}{c}{2--11\%} \\
\hline \noalign{\smallskip}
\end{tabular}
\label{tab:INTEGRAL_AOSummary}
\end{table}

\section{\textbf{Discussion}}

It is interesting to compare our results with those from other missions and facilities. XMM-Newton and \ac{HST} are the obvious missions for comparison as they are similarly long-lived as INTEGRAL and both have much higher average numbers of proposals in response to each AO. There have also been investigations of the outcomes of the observing selection processes for \ac{ESO}, Canadian facilities and the \ac{NRAO} facilities, which are also discussed in Sect.~\ref{subsec:XMMDiscussion}.

For XMM-Newton the difference observed in the proposal selection process of 5--15\% in favour of male PIs is comparable to that observed for INTEGRAL of 2--11\%. Both of these ranges are small in comparison to those reported for some \ac{ESO}, and \ac{NRAO} facilities~\cite{2016Patat}, \cite{2016Lonsdale}, \cite{2020Carpenter} and for \ac{HST} prior to the implementation of dual anonymous reviewing. \cite{2014Reid} reports on a study of gender-based systematic trends in the \ac{HST} proposal review process for Cycles 11 through 21 (2001 to 2013). From this study of the outcomes of 9400 proposals  there appears to be a similar trend to that seen here in that male PIs are more likely to succeed in achieving a successful \ac{HST} proposal (23.5\% success rate) than female PIs (18.1\% success rate). This is a difference of 30\% in favour of proposals led by male PIs. With dual anonymous reviewing the names of the reviewers and the investigators are made known to each other only after the review has been completed. Once dual-anonymous reviewing was implemented for \ac{HST} the male and female PI proposal acceptance difference falls to 9.3\% \cite{2023Reid}, comparable to that seen with INTEGRAL and XMM-Newton.

An investigation of the time allocation process at \ac{ESO} is reported in \cite{2016Patat}. This covers an interval of 8 years and involved about 3000 \ac{PI}s. Female \ac{PI}s were found to have a significantly lower chance of being awarded a top rank compared to male \ac{PI}s with a male/female ratio of 1.39 $\pm$ 0.05. The paper suggests that the principal explanation for the difference may be due to the average higher seniority of the male \ac{PI}s assuming that more senior scientists of both genders write better proposals and thus succeed more often at obtaining \ac{ESO} telescope time. However, no attempt was made to quantify if this effect can account for the observed differences. Nevertheless, \cite{2016Patat} concludes that the \ac{ESO} review process itself introduces extra gender differences.

An investigation of the gender-related systematics in the proposal review processes for the four facilities operated by \ac{NRAO}: the \ac{JVLA}, the \ac{VLBA}, the \ac{GBT} and \ac{ALMA} are reported in \cite{2016Lonsdale} and \cite{2020Carpenter}. Similarly to \ac{HST}, before the introduction of double-anonymous reviewing, and similarly as well to the \ac{ESO} study there are significant gender-related effects in the proposal rankings in favour of male \ac{PI}s compared to female \ac{PI}s for all four facilities with varying degrees of confidence reflecting the different number of proposals.

In summary, the difference observed in the INTEGRAL proposal selection process of 2--11\% in favour of male \ac{PI}s is smaller than those reported for some \ac{ESO}, and \ac{NRAO} facilities~\cite{2016Patat}, \cite{2016Lonsdale} and for \ac{HST} prior to the implementation of dual anonymous reviewing~\cite{2014Reid}. Once dual-anonymous reviewing was implemented for \ac{HST} the male and female PI proposal acceptance difference falls to 9.3\% \cite{2023Reid}, comparable to that seen here with INTEGRAL of 2--11\%. This suggests that the INTEGRAL proposal selection processes produces outcomes consistent with, or at least close to, the best available science. It does not however provide gender parity and it would be useful to investigate reasons for this. The acceptance rates of male and female proposers for different geographical regions could be examined. This could reveal that regional differences play a role, perhaps related to language skills. It would also be useful to investigate the uncertainties in the proposal selection process. This could be accomplished by comparing the results of different representative \ac{TAC}s evaluating the same set of proposals. Finally, it would be interesting to examine the the publication rates of early and late career scientists who have been awarded INTEGRAL observing time. Are late career scientists better at exploiting their observations and so produce relatively more publications and citations than their early career colleagues?  More generally, what is the effect on the science return of a mission if the observations are dominated by late or early career scientists. This is a complex issue as e.g., a late career scientist may have made substantial contributions to a proposal from an early career colleague, so improving its chances of success.

\noindent
\clearpage
\textbf{Acknowledgements.}
\small{We thank the INTEGRAL \ac{SOC} and \ac{ISDC} staff for their support.}

\clearpage
\printbibliography[keyword=INTEGRAL,heading=subbibliography,title={\textbf{References}}]

\clearpage

\section*{\large{Acronym List}}
\addcontentsline{toc}{section}{\large{Acronym List}}

\begin{acronym}[MPCCCCCC] 

\acro{ADS}{Astrophysics Data Service}
\acro{AO}{Announcement of Opportunity}
\acro{ALMA}{Atacama Large Millimeter/sub-millimeter Array}
\acro{CP}{Core Programme}
\acro{D-SCI}{ESA Director of Science}
\acro{ESA}{European Space Agency}
\acro{ESAC}{European Space Astronomy Centre}
\acro{ESO}{European Southern Observatory}
\acro{ESOC}{European Space Operations Centre}
\acro{ESTEC}{European Science and Technology Centre}
\acro{FOV}{Field of View}
\acro{FRB}{Fast Radio Burst}
\acro{FWHM}{Full-Width at Half Maximum}
\acro{GBT}{Green Bank Telescope}
\acro{GP}{General Programme}
\acro{GRB}{Gamma-ray Burst}
\acro{GT}{Guaranteed Time}
\acro{GW}{Gravitional Wave}
\acro{HST}{Hubble Space Telescope}
\acro{IREM}{INTEGRAL Radiation Environment Monitor}
\acro{ISDC}{INTEGRAL Science Data Centre}
\acro{ISWT}{INTEGRAL Science Working Team}
\acro{IBIS}{Imager On-Board the INTEGRAL Spacecraft}
\acro{INTEGRAL}{INTErnational Gamma-Ray Astrophysics Laboratory}
\acro{ISGRI}{INTEGRAL Soft Gamma-Ray Imager}
\acro{ISOC}{INTEGRAL Science Operations Centre}
\acro{IUG}{INTEGRAL Users Group}
\acro{JEM-X}{Joint European X-ray Monitor}
\acro{JVLA}{Jansky Very Large Array}
\acro{keV}{Kilo Electron Volt}
\acro{kpc}{Kilo parsec}
\acro{MeV}{Million Electron Volt}
\acro{MOC}{Mission Operations Centre}
\acro{Msec}{Millions of Seconds}
\acro{NASA}{National Aeronautics and Space Administration}
\acro{NRAO}{National Radio Astronomy Observatory}
\acro{ObsID}{Observation Identifier}
\acro{OMC}{Optical Monitor Camera}
\acro{OSA}{Off-line Scientific Analysis}
\acro{PhD}{Doctor of Philosophy Degree}
\acro{PI}{Principal Investigator}
\acro{PICsIT}{Pixellated Imaging Caesium iodide Telescope}
\acro{SOC}{Science Operations Centre}
\acro{SPI}{Spectrometer on INTEGRAL}
\acro{TAC}{Time Allocation Committee}
\acro{ToO}{Target of Opportunity}
\acro{VLBA}{Very Long Baseline Array}
\acro{XUG}{XMM-Newton Users' Group}

\end{acronym}

\end{document}